\documentclass[a4,12pt]{ozu-thesis}

\usepackage{amsmath,amssymb,latexsym,float,epsfig}
\usepackage{etoolbox}
\usepackage{stfloats}
\usepackage{graphicx}
\usepackage{graphics}
\usepackage[noadjust]{cite}
\usepackage{subcaption}
\usepackage{flushend}
\usepackage{color}
\usepackage{textcmds}
\usepackage{multirow}
\usepackage{physics}
\usepackage{tikz}
\usepackage{mathtools}
\usetikzlibrary{arrows.meta}
\usepackage{nicefrac} 
\usepackage{dirtytalk}
\usepackage{savesym}
    \savesymbol{pdfbookmark}
    \usepackage{hyperref}
    \restoresymbol{HR}{pdfbookmark}
\hypersetup{
    colorlinks=true,
    linkcolor=blue,
    filecolor=magenta,      
    urlcolor=cyan,
}

\title{Performance Analysis of Quantum Key Distribution in Underwater Channels} 
\author{Amir Hossein Fahim Raouf}
\department{Department of Electrical and Electronics Engineering}

\degree{Master of Science}
\copyrightyear{2021}
\submitdate{July 2021} 
                         
\approveddate{27 July 2021}

\begin{document}

\bibliographystyle{ieeetr}

\begin{preliminary}

\begin{dedication}
\null\vfil
{\large
\begin{center}
To whom who fight for freedom, humanitarian and human rights, and dismantling racism
\end{center}}
\vfil\null
\end{dedication}
\begin{abstract}
The current literature on quantum key distribution (QKD) is mainly limited to the transmissions over fiber optic, atmospheric or satellite links and are not directly applicable to underwater environments with different channel characteristics. Absorption, scattering, and turbulence experienced in underwater channels severely limit the range of quantum communication links. 
In the first part of this thesis, we analyze the quantum bit error rate (QBER) and secret key rate (SKR) performance of the well-known BB84 protocol in underwater channels. As path loss model, we consider a modified version of Beer-Lambert formula which takes into account the effect of scattering. We derive a closed-form expression for the wave structure function to determine the average power transfer over turbulent underwater path and use this to obtain an upper bound on QBER as well as a lower bound on SKR. Based on the derived bounds, we present the performance of BB84 protocol in different water types including clear, coastal and turbid water and under different atmospheric conditions such as clear, hazy and overcast. We further investigate the effect of system parameters such as aperture size and detector field-of-view on QBER and SKR performance metrics.
In the second part of this thesis, as a potential solution to overcome range limitations, we investigate a multi-hop underwater QKD where intermediate nodes between the source and destination nodes help the key distribution. We consider the deployment of passive relays which simply redirect the qubits to the next relay node or the receiver without any measurement. Based on the near-field analysis, we present the performance of relay-assisted QKD scheme in terms of quantum bit error rate and secret key rate in different water types and turbulence conditions. We further investigate the effect of system parameters such as aperture size and detector field-of-view on the performance. Our results demonstrate under what conditions relay-assisted QKD can be beneficial and what end-to-end transmission distances can be supported with a multi-hop underwater QKD system.
In the last part of this thesis, we investigate the fundamental performance limits of decoy BB84 protocol over turbulent underwater channels and provide a comprehensive performance characterization. As path loss model, we consider a modified version of Beer-Lambert formula, which takes into account the effect of scattering. Based on near field analysis, we utilize the wave structure function to determine the average power transfer over turbulent underwater path and use this to obtain a lower bound on key generation rate. Based on this bound, we present the performance of decoy BB84 protocol in different water types (clear and coastal). We further investigate the effect of transmit aperture size and detector field of view.
\end{abstract}
\begin{ozetce}
Kuantum anahtar dağıtımı (QKD) hakkındaki mevcut literatür, esas olarak fiber optik, atmosferik veya uydu bağlantıları üzerinden yapılan aktarımlarla sınırlıdır ve farklı kanal özelliklerine sahip su altı ortamlarına doğrudan uygulanamaz. Sualtı kanallarında yaşanan absorpsiyon, saçılma ve türbülans, kuantum iletişim bağlantılarının aralığını ciddi şekilde sınırlar.
Bu tezin ilk bölümünde, iyi bilinen BB84 protokolünün sualtı kanallarında kuantum bit hata oranı (QBER) ve gizli anahtar oranı (SKR) performansını analiz ediyoruz. Yol kaybı modeli olarak, saçılma etkisini hesaba katan Beer-Lambert formülünün değiştirilmiş bir versiyonunu ele alıyoruz. Türbülanslı su altı yolu üzerindeki ortalama güç transferini belirlemek için dalga yapısı fonksiyonu için kapalı biçimli bir ifade türetiyoruz ve bunu QBER'de bir üst sınır ve SKR'de bir alt sınır elde etmek için kullanıyoruz. Elde edilen sınırlara dayanarak, BB84 protokolünün berrak, kıyı ve bulanık su dahil farklı su türlerinde ve berrak, puslu ve bulutlu gibi farklı atmosferik koşullar altında performansını sunuyoruz. Açıklık boyutu ve dedektör görüş alanı gibi sistem parametrelerinin QBER ve SKR performans ölçütleri üzerindeki etkisini ayrıca araştırıyoruz.
Bu tezin ikinci bölümünde, menzil sınırlamalarının üstesinden gelmek için potansiyel bir çözüm olarak, kaynak ve hedef düğümler arasındaki ara düğümlerin anahtar dağıtımına yardımcı olduğu çok atlamalı bir sualtı QKD'sini araştırıyoruz. Kübitleri herhangi bir ölçüm yapmadan bir sonraki röle düğümüne veya alıcıya yönlendiren pasif rölelerin konuşlandırılmasını düşünüyoruz. Yakın alan analizine dayanarak, farklı su türleri ve türbülans koşullarında kuantum bit hata oranı ve gizli anahtar oranı açısından röle destekli QKD şemasının performansını sunuyoruz. Açıklık boyutu ve dedektör görüş alanı gibi sistem parametrelerinin performans üzerindeki etkisini ayrıca araştırıyoruz. Sonuçlarımız, röle destekli QKD'nin hangi koşullar altında faydalı olabileceğini ve çok sekmeli bir sualtı QKD sistemi ile hangi uçtan uca iletim mesafelerinin desteklenebileceğini göstermektedir.
Bu tezin son bölümünde, çalkantılı su altı kanalları üzerinde yem BB84 protokolünün temel performans sınırlarını araştırıyoruz ve kapsamlı bir performans karakterizasyonu sağlıyoruz. Yol kaybı modeli olarak, saçılma etkisini hesaba katan Beer-Lambert formülünün değiştirilmiş bir versiyonunu ele alıyoruz. Yakın alan analizine dayanarak, türbülanslı su altı yolu üzerindeki ortalama güç transferini belirlemek için dalga yapısı fonksiyonunu kullanıyoruz ve bunu anahtar üretim hızında daha düşük bir sınır elde etmek için kullanıyoruz. Bu sınıra dayanarak, yem BB84 protokolünün farklı su türlerinde (temiz ve kıyı) performansını sunuyoruz. İletim açıklığı boyutunun ve dedektör görüş alanının etkisini daha da araştırıyoruz.
\end{ozetce}
\begin{acknowledgements}

I would first like to express my deepest appreciation to my supervisor, Professor Murat Uysal, whose expertise was invaluable in formulating the research questions and methodology. Your insightful feedback pushed me to sharpen my thinking and brought my work to a higher level.

I would also like to sincerely acknowledge the members of my thesis committee,
Dr. Kadir Durak and Dr. Bahattin Karakaya for their time serving on my committee and
carefully reviewing my thesis. 
I would like to extend my sincere thanks to Dr. Majid Safari for his treasured support which was really influential in shaping  my research interests and critiquing my results.
I would like to recognize the assistance, help and effort that I received from the Communication Theory and Technologies (CT\&T) Research Group and my friends.

Last, but not least, I owe my deepest gratitude to my dear parents Afsaneh and Masoud, my sister Sara, and my brother Shayan  for their wise counsel and sympathetic ear. You are always there for me regardless of the situation I may find myself in.

\end{acknowledgements}

\contents
%

\end{preliminary}
\chapter{Introduction}

\section{Motivation}
Underwater sensor networks (USNs) \cite{p1_1} have emerged to provide an integrated system for the surveillance of critical maritime zones and infrastructures, see e.g., \cite{p1_2, p1_3} for some commercially available solutions. With high density node deployment, USNs bring improved agility, resilience and fault tolerance. In addition, they introduce advantages including a higher probability of detection and classification, lower false alarm rate, and more accuracy in the localization of threats. Despite the increasing deployment of USNs and growing relevant literature, cyber security aspects of USNs have received relatively low attention. Particularly for maritime applications such as the surveillance of critical infrastructure and border protection, secure communication is the key to ensure the confidentiality, integrity and authentication of the transmitted information. Some countermeasures for cyber-attacks have been investigated for underwater wireless networks \cite{p1_4, p1_5, p1_6}. For example, a simple symmetric cryptosystem with CipherText Stealing technique is used in \cite{p1_4}. An energy-efficient secure message authentication protocol with the standard encryption techniques is employed in \cite{p1_5}. Similarly, the literature on underwater wireless networks, see e.g., the survey in \cite{p1_6} and references therein, focus only on conventional cryptosystems.

Today's cryptosystems such as widely deployed RSA and elliptic curve-based schemes build upon the formulation of some intractable computational problems. They are able to offer only computational security within the limitations of conventional computing power. 
Recent advances in the quantum computing towards the so-called quantum supremacy have the potential to eventually break such classical cryptosystems \cite{p2_1, p2_2}. Unlike conventional counterparts, quantum cryptography builds upon the laws of quantum mechanics and provides a radically different solution for key distribution promising unconditional security \cite{p2_3}. 

\section{Quantum Key Distribution}
Quantum key distribution (QKD) is used to produce a shared random secret key known only to sender and receiver parties. This key can then be used to encrypt a message, which is transmitted over a standard communication (acoustic, optical, radio, etc.) channel. 
 QKD protocols are based on the concept of the no-cloning theorem \cite{clone}, Heisenberg uncertainty principle \cite{h1} and entanglement property. 
 The main preliminary concepts for better understanding of QKD protocols are briefly summarized as follows.
 
 \subsection{Quantum Bits} 
\begin{figure}
\centering
\begin{tikzpicture}[line cap=round, line join=round, >=Triangle]
  \clip(-2.19,-2.49) rectangle (2.66,2.58);
  \draw [shift={(0,0)}, lightgray, fill, fill opacity=0.1] (0,0) -- (56.7:0.4) arc (56.7:90.:0.4) -- cycle;
  \draw [shift={(0,0)}, lightgray, fill, fill opacity=0.1] (0,0) -- (-135.7:0.4) arc (-135.7:-33.2:0.4) -- cycle;
  \draw(0,0) circle (2cm);
  \draw [rotate around={0.:(0.,0.)},dash pattern=on 3pt off 3pt] (0,0) ellipse (2cm and 0.9cm);
  \draw (0,0)-- (0.70,1.07);
  \draw [->] (0,0) -- (0,2);
  \draw [->] (0,0) -- (-0.81,-0.79);
  \draw [->] (0,0) -- (2,0);
  \draw [dotted] (0.7,1)-- (0.7,-0.46);
  \draw [dotted] (0,0)-- (0.7,-0.46);
  \draw (-0.08,-0.3) node[anchor=north west] {$\varphi$};
  \draw (0.01,0.9) node[anchor=north west] {$\theta$};
  \draw (-1.01,-0.72) node[anchor=north west] {$\mathbf {{x}}$};
  \draw (2.07,0.3) node[anchor=north west] {$\mathbf {{y}}$};
  \draw (-0.5,2.6) node[anchor=north west] {$\mathbf {{z}=\ket{1}}$};
  \draw (-0.4,-2) node[anchor=north west] {$\mathbf {\ket{0}}$};
  \draw (0.4,1.65) node[anchor=north west] {$|\psi\rangle$};
  \scriptsize
  \draw [fill] (0,0) circle (1.5pt);
  \draw [fill] (0.7,1.1) circle (0.5pt);
\end{tikzpicture}
\caption{The Bloch sphere representation of a qubit}\label{fig:bol}
\end{figure}
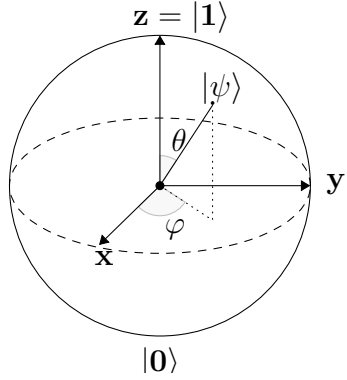

In the information theory concept, the bits are the basic term in classical information where the states of a bit can be either 0 or 1. However, quantum mechanics allows the the quantum bit (qubit) to be in a superposition of both states simultaneously. Let $\ket{1}$ and $\ket{0}$ denotes the two possible states of qubit where $\ket{}$ is called the Dirac notation and is the standard notation for states in quantum mechanics. A general sate for a qubit can be expressed as
$$\ket{\psi} = \alpha  \ket{0} +  \beta \ket{1}
$$
where $\alpha, \beta \in \mathbb{C}$. These factors are complex numbers and the state of a qubit
can be expressed as a vector in a two-dimensional complex vector space $\mathbb{C}^2$ (also known as Hilbert space).
The qubit state is a unit vector, and the normalization property is satisfied as follows
$$
|\alpha|^2 + |\beta|^2 = 1
$$
The state of qubit can also be written as
$$
\ket{\psi} = \cos{\frac{\theta}{2}}   \ket{0} + e^{i\phi}\sin{\frac{\theta}{2}}  \ket{1}
$$
where $\theta$ and $\phi$ are real numbers and define a point on a sphere called the Bloch sphere (see Fig. \ref{fig:bol}).

\subsection{No-cloning Theorem }
The no-cloning theorem states that qubit cannot be copied or amplified without disturbing them. This mechanism enabling the QKD system to detect the presence of eavesdropper (Eve) by using the error parameter measurement that appear during the transmission process of photons from the transmitter to the receiver.

\subsection{Heisenberg Uncertainty Principle}
The Heisenberg uncertainty principle states that certain pairs of physical properties are linked and complementary in the sense that measuring one property will prevent simultaneously knowing of other property and destroys it.

\section{QKD protocols}
QKD protocols can be classified into discrete variable QKD and continuous variable QKD protocol based on the dimension of the source code space. The discrete variable QKD can be further divided into two categories based on entanglement of the light source; namely, prepare and measure (PaM) protocols and Entanglement-based (EB) protocols. 
Prepare-and-measure-based QKD protocols are based on preparing a qubit state and then sent to the other party. 
Later, entanglement-based protocols have emerged in which the two parties share a joint state and perform measurements on that. Although entanglement protocols offer an additional level of security as the quantum source does not have to be trusted, PaM protocols have been more popular mainly due to their simplicity. Table \ref{table_qkd} summarize the main powerful QKD protocols.

The first QKD protocol proposed by Bennett and Brassard, today widely known as \textbf{BB84} \cite{p1_8}. The BB84 protocol is based on the concept of quantum coding proposed by Wiesner's \cite{w1}. In BB84, information is encoded in orthogonal
quantum states. The transmitter (i.e., Alice) prepares a qubit by choosing randomly between two linear polarization bases $\oplus$ or $\otimes$ for every bit she wants to send. She selects a random bit value \say{0} or \say{1} for each chosen base, using the following polarization rule

$$
  \oplus =
    \begin{cases}
      0^{\circ} & \text{if \say{0}  was chosen}\\
      +90^{\circ} & \text{if \say{1}  was chosen}
    \end{cases}       
$$
$$
  \otimes =
    \begin{cases}
      -45^{\circ} & \text{if \say{0}  was chosen}\\
      +45^{\circ} & \text{if \say{1}  was chosen}
    \end{cases}       
$$

The receiver (i.e., Bob) measures the possible arrival qubit in either $\oplus$ or $\otimes$ bases. 
For a given qubit, the same result will be revealed for both Alice and Bob if they measure on the same basis.
If Alice send a qubit in a $\oplus$ and Bob measure it in $\otimes$, then there is 50-50\% chance
of getting $-45^{\circ}$ or $+45^{\circ}$. 
Similarly, If Alice send a qubit in $\otimes$ and Bob measure it in $\oplus$, then there is 50–50\% chance of getting
 $0^{\circ}$ or $+90^{\circ}$.
As a result, Alice and Bob construct the secure key based on the qubits received at the \say{sift} events. Sift events correspond to the bit intervals in which exactly one of the detectors registers a count and both Alice and Bob have chosen the same basis. Alice and Bob can recognize the sift events by transferring information over a public classical communication channel. 
In practice, in addition to Eve's potential intervention, the sifted key contains errors caused by path loss, turbulence, and noise in Bob's detectors. Error correction is employed to reduce the effect of such channel
imperfections. Further, in order to minimize the information compromised to potential Eve, privacy amplification is employed. This process can be accomplished by deliberately discarding a number of bits in the generated key to decrease the information acquired by the potential Eve. 

Bennett, Brassard and Mermin \cite{bbm92} proposed an entanglement version of BB84 called \textbf{BBM92}. 
In BBM92, both Alice and Bob measure the arriving photons from a central source.
The results for Alice and Bob are perfectly correlated when they use the same measurement basis. If
Alice and Bob choose a different basis (e.g., Alice choose $\otimes$ and Bob choose $\oplus$), then their results will
not be correlated.
Waks \textit{et al.} \cite{10} studied the security proof for BBM92 protocol with realistic and untrusted source against individual attacks. Their results showed that the average collision probability of BBM92 is the same as BB84, while BBM92 outperforms BB84 in the terms of communication rate.
The main advantage of BBM92 protocol is that Alice and Bob can notice any malicious control by Eve to the source. 
It should be noted that a trusted central source for generating entangled photon is not required.

In 1992, Bennett proposed a novel protocol called \textbf{B92} using
one non-orthogonal basis. 
In \cite{76, 77}, Tamaki \textit{et al.} investigated the security of B92 employing a single-photon source. 
Further and Koashi \cite{78} studied the B92 implementation employing strong phase-reference coherent light. 
It has also been shown that B92 protocol performs better in term of eavesdropper detection as compared to BB84 \cite{79}.

Ekert proposed a novel entanglement based protocol named \textbf{E91} which employs the generalized Bell's theorem for testing of eavesdropping.  
In this protocol, a sequence of entangled pairs of qubits from a central source transmitted to Alice and Bob. 
Two compatible cases can be occurred as follows:
\begin{itemize}
    \item If Alice (Bob) measures spin up, then Bob (Alice) collapses into a spin down.
    \item If Alice (Bob) measures spin down, then Bob (Alice) collapses into spin up.
\end{itemize}

It is worth mentioning that the order of measurement is not important, i.e., if Alice (Bob) measures the first pair, then Bob (Alice) will collapse respectively.
In E91 protocol, Alice's and Bob's results are \textit{perfectly} correlated or not correlated which result in identifying the
Eavesdropper.
In original E91 protocol, Ekert proposed three bases for Alice ($0^{\circ}$, $45^{\circ}$, and $90^{\circ}$)
and Bob ($45^{\circ}$, $90^{\circ}$, and $135^{\circ}$). In this scheme, the chance of measuring in the compatible bases are $\nicefrac{1}{3}$.
Similar to BB84, Alice and Bob publicly announce their bases and discard the incompatible bases. 
In \cite{85}, Ilic studied different concepts of error correction, privacy amplification and violation of Bell's theorem in E91 protocol.
A simplified version of E91 is proposed by Acin \textit{et al.} \cite{89} where it considers three bases on one side and two bases on the other side. 
An experimental demonstration of Acin \textit{et al.} protocol through 20 KM fiber link is further reported in \cite{91}.

Bruß \cite{6states} generalized the BB84 protocol and proposed \textbf{six-state} protocol employing three conjugate bases. These six states are pointing towards positive and negative of x-axis, y-axis
and z-axis of the Bloch sphere. 
Bruß \cite{6states} also proved that six-state protocol are more secure than BB84 protocol.
An unconditional security of the six-state protocol is investigated in \cite{95}.

Scarani \textit{et al.} \cite{sarg} proposed a novel protocol named \textbf{SARG04} which is robust against photon number splitting (PNS) attack\footnote{In case of multi-photon emission an Eve is able to launch a PNS attack; an adversary stops all single-photon signals and splits multi-photon signals, keeping one copy herself and resending the rest to the legitimate receiver.} with weak pulses. 
This protocol utilizes two non-orthogonal quantum states similar to B92 protocol.
As opposed to BB84 protocol, Alice never announces her basis to Bob in SARG04 protocol.
An entangled version of SARG04 is proposed by Branciard \textit{et al.} in \cite{102}. They showed that SARG04 outperforms BB84 in terms of maximal achievable distance for a wider class of Eve's attacks.
Further,  Koashi \cite{103} generalized the SARG04 protocol to $n$ quantum state protocol.

The \textbf{decoy state} concept was first introduced by Hwang \cite{p4_15} and the first complete security proof of the decoy-method was given in 2005 by Lo \textit{et al.} \cite{p4_16} considering an infinite amount of intensities. 
The decoy method is able to combat PNS attack with utilizing different pulse intensities of decoy states dispersed randomly within the signal pulses \cite{p4_14}.  Decoy state protocol utilizes imperfect single-photon sources e.g., weak coherent state source as the transmitter. 
In such protocols, the transmitter sends the decoy state pulse sequence (contains no useful information) accompanying the single photon pulse sequence. Since Eve cannot distinguish whether a photon state is from a signal or a decoy, her attempts on photon number splitting attack leads to a variation on the expected yield of signal and decoy states. 
In \cite{313}, Liao \textit{et al.} reported the implementation of decoy-sate BB84 for a low-earth satellite for a link distance up to 1200 KM.
Further, in \cite{p2_15}, they implemented the decoy-state BB84 between China to Austria with low-earth orbit satellite. 

 In continuous variable QKD protocol, the information is transmitted in the form of light rather than single photon. The main advantage of such protocols is that the measurement technique (i.e., Homodyne detection) can work at a high rate.
 The concept of continuous variable quantum key distribution is independently introduced by Ralph \cite{g1} and Reid \cite{20}. 
A continuous analogue of B84 protocol employing squeezed states of light and Homodyne detection mechanism is introduced in \cite{21}. 
Leverrier and Grangier \cite{29} proposed two continuous variable QKD protocols with discrete modulation utilizing two and four coherent states.
More recently, Papanastasiou and Pirandola \cite{30} proposed a continuous variable QKD protocol employing discrete-alphabet
encoding.


\begin{table}
\captionsetup{justification=centering}
\caption{Summary of the most famous QKD protocols}
\label{table_qkd}
\begin{center}
\scalebox{0.95}{
\begin{tabular}{llll}
\hline
\multicolumn{1}{|l|}{Signal Type} & \multicolumn{1}{l|}{Protocol Type} & \multicolumn{1}{l|}{Name} & \multicolumn{1}{l|}{Year} \\ \hline
\multicolumn{1}{|c|}{\multirow{7}{*}{Discrete Variable QKD}} & \multicolumn{1}{l|}{\multirow{5}{*}{Prepare and Measure Protocols}} & \multicolumn{1}{l|}{BB84 \cite{p1_8}} & \multicolumn{1}{l|}{1984} \\ 
\cline{3-4} 
\multicolumn{1}{|c|}{} & \multicolumn{1}{l|}{} & \multicolumn{1}{l|}{B92 \cite{b92}} & \multicolumn{1}{l|}{1992} \\
\cline{3-4} 
\multicolumn{1}{|c|}{} & \multicolumn{1}{l|}{} & \multicolumn{1}{l|}{Six-State \cite{6states}} & \multicolumn{1}{l|}{1998} \\
\cline{3-4} 
\multicolumn{1}{|c|}{} & \multicolumn{1}{l|}{} & \multicolumn{1}{l|}{Decoy-state \cite{p4_15}} & \multicolumn{1}{l|}{2003} \\ \cline{3-4} 
\multicolumn{1}{|c|}{} & \multicolumn{1}{l|}{} & \multicolumn{1}{l|}{SARG04 \cite{sarg}} & \multicolumn{1}{l|}{2004} \\ \cline{2-4} 
\multicolumn{1}{|c|}{} & \multicolumn{1}{l|}{\multirow{2}{*}{Entanglement Based}} & \multicolumn{1}{l|}{E91 \cite{p1_9}} & \multicolumn{1}{l|}{1991} \\ \cline{3-4} 
\multicolumn{1}{|c|}{} & \multicolumn{1}{l|}{} & \multicolumn{1}{l|}{BBM92 \cite{bbm92}} & \multicolumn{1}{l|}{1992} \\ \hline
\multicolumn{1}{|l|}{\multirow{2}{*}{Continuous Variable QKD}} & \multicolumn{2}{l|}{Gaussian Modulation \cite{g1}} & \multicolumn{1}{l|}{2002} \\ \cline{2-4} 
\multicolumn{1}{|l|}{} & \multicolumn{2}{l|}{Discrete Modulation \cite{29}} & \multicolumn{1}{l|}{2011} \\ \hline
\end{tabular}
}
\end{center}
\end{table}

Most of QKD protocols are limited to binary signal format, i.e., qubits, which are two-level quantum systems. To take advantage of the higher dimensionality, orbital angular momentum (OAM) is further used to design QKD systems \cite{p1_13, p1_14}. In these systems, the encoded quantum states belong to a higher dimensional Hilbert space where qudits rather than qubits are used. The feasibility of various QKD protocols has been further demonstrated through successful experiments for different transmission ranges and data rates, see e.g., a recent survey \cite{p1_15} and references therein.

\section{Challenges in Quantum Key Distribution}
The main practical challenges in quantum key distribution systems can be summarized as follows.

\begin{itemize}

    \item \textbf{Photon Source:}
The security of using BB84 is bound to single photon transmitter. 
In case of multi-photon emission an Eve is able to launch a photon number splitting attack (PNS). Current implementations employ faint laser pulses where most time slots are empty, a few have single photons and very few of them have more than a single photon.

    \item \textbf{Communication Distance:}
  Due to the detector noise and channel losses, the range of current quantum key
distribution systems is limited. In addition, this limitation
can be associated with the fact that BB84 and other single photon based
protocols operate with an average number of photons per pulse much less than 1.

    \item \textbf{Cost and Robustness:}
The cost and robustness features of QKD systems alongside high performance are indispensable for real world applications.

\end{itemize}

\section{Literature Review}
The current literature on QKD is mainly limited to the transmissions over fiber optic, atmospheric or satellite links and are not directly applicable to underwater environments with different channel characteristics. There have been only some recent efforts on underwater QKD \cite{p1_16, p1_17, p1_18, p1_19, p1_20, p1_21, p1_22, p1_23, p1_24, p1_25}. For example, in \cite{p1_16} and \cite{p1_17}, based on the well-known Beer-Lambert path loss model, the maximum secure communication distance for BB84 protocol in underwater environments was derived to achieve a desired level of quantum bit error rate (QBER) and the secret key rate (SKR). In \cite{p1_18} and \cite{p1_19}, Monte Carlo simulations were conducted to determine the propagation characteristics of polarized photons in the underwater channel. Using these simulated underwater channels, the QBER performance of BB84 QKD protocol was computed. Some experimental demonstrations of underwater QKD were further reported in \cite{p1_20, p1_21, p1_22} using aquariums and water tanks. Specifically, in \cite{p1_20}, Bouchard \textit{et al.} implemented 2- and 3-dimensional BB84 protocols and reported results for a transmission distance of $3$ meters. In another experiment over a $2.37$ m distance \cite{p1_21}, Zhao \textit{et al.} reported that QBER less than 3.5\% can be achieved for different water types with extinction coefficients up to 0.7  $\rm{m^{-1}}$. In \cite{p1_22}, Hu \textit{et al.} used a semi-open water tank and experimentally verified the feasibility of underwater QKD over a transmission distance of $55$ meters.

The above theoretical and experimental works consider only the path loss, but ignore the effects of turbulence. In practical scenarios, rapid changes in the refractive index commonly caused by ocean currents induce sudden variations in the water temperature and pressure. This turbulence results in fluctuations of the optical signal known as fading. The effects of underwater turbulence on QKD systems were addressed only recently in \cite{p1_23, p1_24, p1_25}. In \cite{p1_23}, Bouchard \textit{et al.} experimentally investigated the effect of turbulence on an OAM-based QKD system in an outdoor swimming pool exposed to temperatures in the range of $17^{\circ}$ C - $27^{\circ}$ C and demonstrated performance degradations due to turbulence. In {\cite{p1_24}}, Hufnagel \textit{et al.} carried out an experiment in Ottawa River for a 5.5 meter quantum link and quantified the effect of turbulence on optical beams with different polarization states and spatial modes. In \cite{p1_25}, Gariano and Djordjevic adopted the split step beam propagation simulation method to model the effect of oceanic turbulence and used this model to calculate the SKR of BB84 protocol.

The above experimental and theoretical studies point out that performance degradation due to absorption, scattering, and turbulence experienced in underwater channels severely limit the range of quantum communication links. As a potential solution to overcome range limitations, relay-assisted QKD is introduced in the literature. The concept of relay-assisted QKD was earlier studied for atmospheric, fiber and satellite links \cite{p2_15, p2_16, p2_17}, however those results are not directly applicable to underwater communications which features inherent differences. Underwater optical communication suffers from severe absorption and scattering due to the inevitable photon interactions with the water molecules and other particles in solution and suspension in water. The maximum transmission distance depends on the type of water and concentration of dissolved particles therein. Furthermore, the operation wavelength is typically in the blue and green spectrum which is distinct from those of free space and fiber optic communications. Therefore, it remains an open question to find out if relay-assisted transmission is beneficial in such a harsh propagation environment and what end-to-end transmission distances can be supported with a multi-hop underwater QKD system.


\section{Thesis Outlines}
The main contributions and the results of this thesis which have been also reported in \cite{p2_14, raouf2020multi,FahimRaouf:20, Raouf:22} can be categorized in three parts as follows.

In the first part of this thesis, we analyze the QBER and SKR performance of the well-known BB84 protocol in underwater channels. As path loss model, we consider a modified version of Beer-Lambert formula which takes into account the effect of scattering. We derive a closed-form expression for the wave structure function to determine the average power transfer over turbulent underwater path and use this to obtain an upper bound on QBER as well as a lower bound on SKR. Based on the derived bounds, we present the performance of BB84 protocol in different water types including clear, coastal and turbid water and under different atmospheric conditions such as clear, hazy and overcast. We further investigate the effect of system parameters such as aperture size and detector field-of-view on QBER and SKR performance metrics.

In the second part of this thesis, as a potential solution to overcome range limitations, we investigate a multi-hop underwater QKD where intermediate nodes between the source and destination nodes help the key distribution. We consider the deployment of passive relays which simply redirect the qubits to the next relay node or the receiver without any measurement. Based on the near-field analysis, we present the performance of relay-assisted QKD scheme in terms of quantum bit error rate and secret key rate in different water types and turbulence conditions. We further investigate the effect of system parameters such as aperture size and detector field-of-view on the performance. Our results demonstrate under what conditions relay-assisted QKD can be beneficial and what end-to-end transmission distances can be supported with a multi-hop underwater QKD system.


In the last part of this thesis, we investigate the fundamental performance limits of decoy BB84 protocol over turbulent underwater channels and provide a comprehensive performance characterization. As path loss model, we consider a modified version of Beer-Lambert formula, which takes into account the effect of scattering. Based on near field analysis, we utilize the wave structure function to determine the average power transfer over turbulent underwater path and use this to obtain a lower bound on key generation rate. Based on this bound, we present the performance of decoy BB84 protocol in different water types (clear and coastal). We further investigate the effect of transmit aperture size and detector field of view.

\section{Organization}

The rest of thesis is organized as follows. In Chapter \ref{ch:QKD}, we investigate the fundamental performance limits of BB84 protocol over turbulent underwater channels and provide a comprehensive performance characterization. In Chapter \ref{ch:relay}, we consider a multi-hop underwater QKD system where relay nodes are utilized along the path connecting two legitimate parties. 
In Chapter \ref{ch:decoy}, we investigate the fundamental performance limits of decoy BB84 protocol over turbulent underwater channels and provide a comprehensive performance characterization. Finally, Chapter \ref{ch:conc} concludes and summarizes this thesis.

\chapter{QKD in Underwater Turbulence Channels}\label{ch:QKD}

In this chapter, we investigate the fundamental performance limits of BB84 protocol over turbulent underwater channels and provide a comprehensive performance characterization. As path loss model, we consider a modified version of Beer-Lambert formula, which takes into account the effect of scattering. We derive a closed-form expression for the wave structure function to determine the average power transfer over turbulent underwater path and use this to obtain an upper bound on QBER and a lower bound on SKR. Based on these bounds, we present the performance of BB84 protocol in different water types (clear, coastal and turbid) and different atmospheric conditions (clear, hazy and overcast atmosphere with relative locations of sun and earth at day time). We further investigate the effect of transmit aperture size and detector field-of-view (FOV) on the system performance.

The remainder of this chapter is organized as follows. In Section \ref{sec:P1_sys}, we describe our system model based on BB84 QKD protocol. In Section \ref{sec:P1_Per_Analysis}, we derive the underwater wave structure function and analyze the QBER and SKR in the presence of turbulence. In Section \ref{sec:P1_Results}, we present numerical results to corroborate on the derived expressions.

\vspace{5pt}
\section{System Model}\label{sec:P1_sys}

\begin{figure}[tb]
\centering
\includegraphics[width=0.8\linewidth]{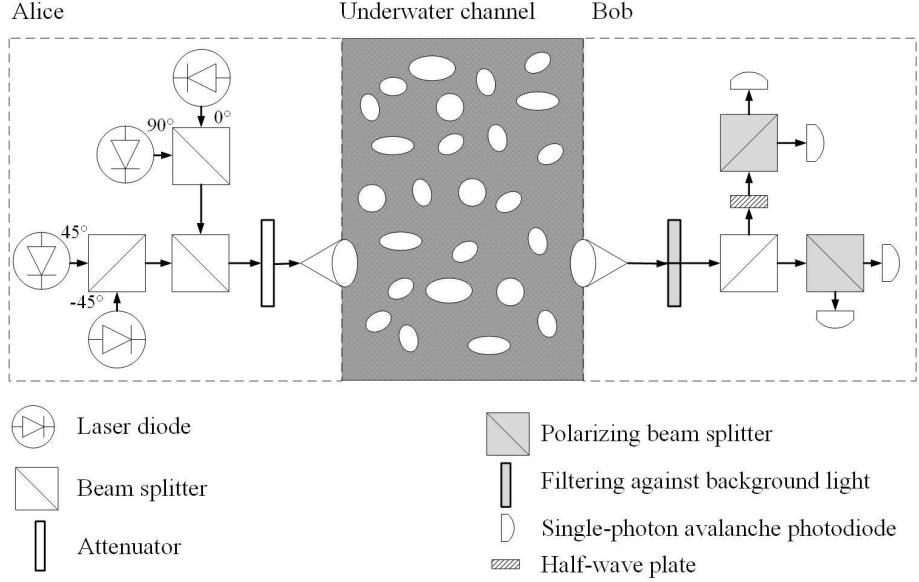}
\caption{Underwater BB84 QKD system under consideration}
\label{fig:1}
\end{figure}

Fig. \ref{fig:1} illustrates a schematic diagram of a typical QKD system which uses BB84 protocol for key distribution. In this protocol, the authorized partners, Alice and Bob, wish to establish a secret key about which no Eve can acquire noteworthy information. Alice prepares a qubit by choosing randomly between two linear polarization bases $\oplus$ or $\otimes$ for every bit she wants to send. She selects a random bit value \say{0} or \say{1} for each chosen base, using the following polarization rule

$$
  \oplus =
    \begin{cases}
      0^{\circ} & \text{if \say{0} was chosen}\\
      +90^{\circ} & \text{if \say{1} was chosen}
    \end{cases}       
$$
$$
  \otimes =
    \begin{cases}
      -45^{\circ} & \text{if \say{0} was chosen}\\
      +45^{\circ} & \text{if \say{1} was chosen}
    \end{cases}       
$$

At the receiver side, a passive 50:50 beam splitter chooses a random basis from the two bases. At the outputs of the beam splitter, two polarization detectors measure the quantum state of the possibly coming photon based on the two different bases. Each of these units includes a polarizing beam splitter (PBS) to decide between two orthogonal polarization states of the corresponding basis and two single-photon avalanche photodiodes (APDs) operating in Geiger-Mode for photon counting. Alice and Bob construct the secure key based on the qubits received at the \say{sift} events. Sift events correspond to the bit intervals in which exactly one of the APDs registers a count and both Alice and Bob have chosen the same basis. Alice and Bob can recognize the sift events by transferring information over a public classical communication channel (in our case underwater optical channel). In practice, in addition to Eve's potential intervention, the sifted key contains errors caused by path loss, turbulence and noise in Bob's detectors. Error correction is employed to reduce the effect of such channel imperfections {\cite{p1_26}}.

Assume that Alice transmits a normalized spatial beam pattern from circular exit pupil and a diameter of $d_1$ with an average photon number of $n_S$ to represent her bit value. Bob collects the light received from Alice with a diameter of $d_2$ in the $z = L$ plane. The effects of diffraction, turbulence and attenuation loss lead to a reduction in Bob's collected photons. In addition, Bob's receiver will collect $n_B$ background photons per polarization on average, and each of his detectors will be subject to an average equivalent dark current photon number of $n_D$. By considering the dark current and irradiance of the environment, the average number of noise photons reaching each Bob's detector can be obtained by \cite{p1_27}
\vspace{-5pt}
\begin{equation}\label{eq:1}
n_N = \frac{n_B}{2} + n_D = I_{dc}\Delta t + \frac{1}{2}\frac{\pi R_d A \Delta t\ensuremath{'} \lambda \Delta \lambda (1 - \cos(\Omega))}{2 h_p c_{light}}
\end{equation}
where $I_{dc}$ is the dark current count rate, $A$  is the receiver aperture area,  $\Omega$  is the FOV of the detector, $h_p$  is Planck's constant, $c_{light}$ is the speed of light, $R_d$ is the irradiance of the environment, $\Delta \lambda$ is the filter spectral width, $\Delta t$ is the bit period and $\Delta t\ensuremath{'}$ is the receiver gate time. It is convenient to write the depth dependence of $R_d(\lambda,z_d)$ as
\vspace{-5pt}
\begin{equation}\label{eq:2}
R_d(\lambda,z_d) = R_d(\lambda,0)\exp{(-k_\infty z_d)}
\end{equation}
where $k_\infty$ is the asymptotic value of the spectral diffuse attenuation coefficient for spectral downwelling plane irradiance \cite{p1_28}. The typical total irradiances at sea level in the visible wavelength band for some atmospheric conditions are provided in \cite{p1_29}.
\vspace{-5pt}
\section{Performance Analysis}\label{sec:P1_Per_Analysis}
In this section, we investigate the performance of the underwater QKD system through the derivation of an upper bound on QBER and a lower bound on SKR.
\subsection{QBER Analysis}
QBER is defined as the error rate in the sifted key. Mathematically speaking, it is given by {\cite{p1_30}}
\vspace*{-0.2cm}
\begin{equation}\label{eq:3}
\text{QBER} = \frac{\text{Pr(error)}}{\text{Pr(sift)}}
\end{equation}
where Pr(error) and Pr(sift) are probabilities of sift and error, respectively. Replacing lower and upper bounds for sift and error probabilities obtained in \cite{p1_31} over a turbulent channel based on the near-field analysis, a lower bound on QBER is obtained as 
\begin{equation}\label{eq:4}
\begin{gathered}
{\rm{QBER}} \le \frac{{{n_N}\left( {1 - \mu  + \mu {e^{ - \eta {n_S}l}}} \right)}}{{\frac{{{n_S}l}}{2}\mu {e^{ - \eta {n_S}l}} + 2{n_N}\left( {1 - \mu  + \mu {e^{ - \eta {n_S}l}}} \right)}}
\end{gathered}
\end{equation}
where $\eta$ is the quantum efficiency of APDs, $l$ is the path loss and $\mu$ is the average power transfer over the turbulent path. Calculation of $l$ and $\mu$ depends on the operation environment and will be discussed in the following for the underwater channel under consideration.

The underwater path loss is a function of attenuation and geometrical losses. For collimated light sources, the geometrical loss is negligible; therefore, the path loss with a laser diode transmitter only depends on the attenuation loss. The attenuation loss is characterized by wavelength-dependent extinction coefficient $\varsigma =\alpha + \beta$ where $\alpha$ and $\beta$ respectively denote absorption and scattering coefficients. Typical values of absorption and scattering coefficients for different water types including clear ocean, coastal water, and turbid harbor can be found in Table \ref{table1} for $\lambda = 532$ $\rm{nm}$, i.e., in the blue-green spectral region \cite{p1_32}. In our work, we utilize the modified version of Beer-Lambert formula proposed in \cite{p1_33}, which takes into account the contribution of scattered lights. The path loss for a transmission distance of $L$ can be given as 
\begin{equation}\label{eq:5}
l = \exp \left[ { - \varsigma L{{\left( {\frac{{{d_2}}}{{\theta L}}} \right)}^T}} \right]
\end{equation}
where $\theta$ is full-width transmitter beam divergence angle and $T$ is a correction coefficient \cite{p1_33}.

The average power transfer over the turbulent path is expressed as \cite{p1_31}
\begin{equation}\label{eq:6}
\begin{gathered}
\mu = \frac{8 \sqrt{F}}{\pi} 
\int_0^1 e^{(\frac{-W(d_1 x,L)}{2})} \bigg( \arccos(x) - x \sqrt{1-x^2} \bigg) J_1(4 x \sqrt{F}) dx
\end{gathered}
\end{equation}
where $F$ is the Fresnel number given by $F = \big( (\pi d_1 d_2)/4 \lambda L \big)^2$ and $J_1 (\cdot)$ is the first-order Bessel function of the first kind. Here, $W (\cdot,\cdot)$ is the wave structure function and is related to spatial power spectrum of refractive index, therefore dependent on the operation environment. Let $\rho$ denote the distance between two observation points. For spherical waves, wave structure function can be calculated as \cite{p1_34}
\begin{equation}\label{eq:7}
W(\rho,L) = 8 \pi^2 k^2 L \int_0^1 \int_0^{\infty} [1-J_0(\kappa \zeta \rho)]\Phi(\kappa)\kappa d\kappa d\zeta
\end{equation}
where $J_0 (\cdot)$ is the zero-order Bessel function, $k=2\pi / \lambda$ is the wave number and $\Phi(\kappa)$ is three-dimensional spatial power spectrum of refractive index in turbulent ocean. In the following, we derive a closed-form expression for underwater wave structure function.

Assume that  $\varepsilon$ denotes the dissipation rate of turbulent kinetic energy per unit mass of fluid. Let $\alpha_{th}$ and $\chi_T$ respectively denote the thermal expansion coefficient and the dissipation rate of mean-squared temperature. Furthermore, let $d_r$ denote the eddy diffusivity ratio. Based on the modified Nikishov spectrum \cite{p1_35}, $\Phi(\kappa)$  is given by 
\begin{equation}
\begin{split}
 \Phi (\kappa ) & = 0.18\left( {\frac{{\alpha _{th}^2{\chi _T}}}{{{\omega ^2}}}} \right)\frac{{{{\left( {\varepsilon {\kappa ^{ - 11}}} \right)}^{ - \frac{1}{3}}}}}{\pi }\left[ {1 + g{\kappa ^{\frac{2}{3}}}} \right] \\
 & \! \times \! \!
\left(\!\! {{\omega ^2} \exp{( - a{\kappa ^{\frac{4}{3}}} - b{\kappa ^2})} \!+\! {d_r} \exp{\!(\!\! -\! c{\kappa ^{\frac{4}{3}}} - d{\kappa ^2}\!)\!} \!-\! \omega ({d_r} + 1) \exp{\!( - e{\kappa ^{\frac{4}{3}}}\!\! -\!\! f{\kappa ^2})}}\! \right)
\end{split}
\label{eq:8}
\end{equation}

In \eqref{eq:8}, $a = 1.08 P_T^{-1}\eta_K^{4/3}$, $b=1.692 P_T^{-1}\eta_K^2$, $c = 1.08 P_S^{-1}\eta_K^{4/3}$, $d = 1.692 P_S^{-1} \eta_K^2$, $e = 0.54 P_{TS}^{-1} \eta_K^{4/3}$, $f = 0.846 P_{TS}^{-1} \eta_K^2$ and  $g = 2.35 \eta_K^{2/3}$.  Here, $\eta_K$ is Kolmogorov microscale length and given by  ${\eta _K} = {\left( {{{{\upsilon ^3}} \mathord{\left/
 {\vphantom {{{\upsilon ^3}} \varepsilon }} \right.
 \kern-\nulldelimiterspace} \varepsilon }} \right)^{{1 \mathord{\left/
 {\vphantom {1 4}} \right.
 \kern-\nulldelimiterspace} 4}}}$ with $\upsilon$ referring to the kinematic viscosity. Furthermore, $P_T$ is the Prandtl number of temperature, $P_S$ is the Prandtl number of salinity, $P_{TS}$ is one half of the harmonic mean of  $P_T$ and $P_S$, and  $\omega$ is the relative strength of temperature and salinity fluctuations. By replacing \eqref{eq:8} inside \eqref{eq:7} and expanding the zero-order Bessel function in terms of power series, we can express the wave structure function as 

\begin{equation}
\begin{split}
W\!\!\left(\! {\rho ,L} \!\right)\!\! =\!\! & 1.44\pi {k^2}L\alpha _{th}^2{\chi _T}{\varepsilon ^{\left( { - 1/3} \right)}}\int\limits_0^1 {\sum\limits_{n = 1}^\infty  {\frac{{{{\left( { - 1} \right)}^{n - 1}}{{\left( {\rho \zeta } \right)}^{2n}}}}{{{{\left( {n!} \right)}^2}{2^{2n}}}}d\zeta } } \int\limits_0^\infty  {{\kappa ^{2n - \frac{8}{3}}}} \exp \left( { - a{\kappa ^{\frac{4}{3}}} - b{\kappa ^2}} \right)d\kappa\!\! {\rm{ }}\\
+&\! 1.44\pi \!{k^2}L\left(\!\! {\frac{{\alpha _{th}^2{\chi _T}}}{{{\omega ^2}}}}\!\! \right){d_r}\!{\varepsilon ^{\left( { - 1/3} \right)}}\!\int\limits_0^1\! {\sum\limits_{n = 1}^\infty \!\! {\frac{{{{\left( { - 1} \right)}^{n - 1}}{{\left( {\rho \zeta } \right)}^{2n}}}}{{{{\left( {n!} \right)}^2}{2^{2n}}}}d\zeta \!\!} } \int\limits_0^\infty\!\! \!\! {{\kappa ^{2n - \frac{8}{3}}}\exp \left( { - c{\kappa ^{\frac{4}{3}}} - d{\kappa ^2}}\! \right)\!d\kappa \!\!} \\
-&\!\! 1.44\pi\! {k^2}L\!\left(\!\! {\frac{{\alpha _{th}^2{\chi _T}}}{\omega }}\!\! \right)\!\left(\!\! {{d_r}\! +\! 1}\! \!\!\right){\!\varepsilon ^{\left( { - 1/3} \right)}}\!\!\!\int\limits_0^1 \!{\sum\limits_{n = 1}^\infty \! {\frac{{{{\left( { - 1} \right)}^{n - 1}}{{\left( {\rho \zeta } \right)}^{2n}}}}{{{{\left( {n!} \right)}^2}{2^{2n}}}}d\zeta\!\! } }\! \int\limits_0^\infty\!\!  {{\kappa ^{2n - \frac{8}{3}}}\exp (\! - \!e{\kappa ^{\frac{4}{3}}}\!\! -\! \!f{\kappa ^2})d\kappa\!\! } \\
 +&\! 1.44\pi \!{k^2}\!L\alpha _{th}^2{\chi _T}{\varepsilon ^{\left( { - 1/3} \right)}}g\!\int\limits_0^1\! {\sum\limits_{n = 1}^\infty\!  {\frac{{{{\left( { - 1} \right)}^{n - 1}}{{\left( {\rho \zeta } \right)}^{2n}}}}{{{{\left( {n!} \right)}^2}{2^{2n}}}}d\zeta } } \int\limits_0^\infty  {{\kappa ^{2n - 2}}} \!\exp \left(\! { - a{\kappa ^{\frac{4}{3}}} - b{\kappa ^2}} \!\right)\!d\kappa \!\!\\
+& 1.44\pi {k^2}L\!\!\left(\! \!{\frac{{\alpha _{th}^2{\chi _T}}}{{{\omega ^2}}}} \!\!\right)\!{d_r}{\varepsilon ^{\left( { - 1/3} \right)}}g\!\!\!\int\limits_0^1 {\sum\limits_{n = 1}^\infty \! {\frac{{{{\left( { - 1} \right)}^{n - 1}}{{\left(\! {\rho \zeta }\! \right)}^{2n}}}}{{{{\left( {n!} \right)}^2}{2^{2n}}}}\!d\zeta } }\!\! \int\limits_0^\infty\!  {{\kappa ^{2n - 2}}} \!\exp \left(\! { - c{\kappa ^{\frac{4}{3}}} - d{\kappa ^2}}\!\! \right)\!d\kappa\!\! \\
-&\! 1.44\pi\! {k^2}L\!\!\left(\!\! {\frac{{\alpha _{th}^2{\chi _T}}}{\omega }} \!\!\right)\!\left(\!\! {{d_r}\! +\! 1} \!\right){\varepsilon ^{\left( { - 1/3} \right)}}g\!\!\!\int\limits_0^1 {\sum\limits_{n = 1}^\infty \! \!{\frac{{{{\left( { - 1} \right)}^{n - 1}}{{\left(\! {\rho \zeta }\! \right)}^{2n}}}}{{{{\left( {n!} \right)}^2}{2^{2n}}}}d\zeta \!\!} } \!\int\limits_0^\infty\!\!\!\!  {{\kappa ^{2n - 2}}} \!\exp (\! - e{\kappa ^{\frac{4}{3}}}\!\! -\!\! f{\kappa ^2})d\kappa 
\end{split}
\label{eq:9}
\end{equation}

From \eqref{eq:9}, it can be observed that the first three integrals have similar forms. For the convenience of presentation, consider only the first integral. Using Eq. (4) of \cite{p1_36}, Eq. (11.1) of \cite{p1_37}, Eq. (9.2) of \cite{p1_37}), and Eq. (b) in [Ch. 9, 37], we can define it in terms of the generalized hypergeometric function as
\begin{equation}\label{eq:10}
\begin{split}
 \int\limits_0^1 & {\sum\limits_{n = 1}^\infty  {\frac{{{{\left( { - 1} \right)}^{n - 1}}{{\left( {\zeta \rho } \right)}^{2n}}}}{{{{\left( {n!} \right)}^2}{2^{2n}}}}\int\limits_0^\infty  {{\kappa ^{2n - \frac{8}{3}}}\exp \left( { - a{\kappa ^{\frac{4}{3}}} - b{\kappa ^2}} \right)d\kappa } } d\zeta }  \\
  = & \frac{1}{2}{b^{\frac{5}{6}}}\Gamma \left( { - \frac{5}{6}} \right)\left[ {1 - {}_2{F_2}\left( {\frac{{ - 5}}{6},\frac{1}{2};1,\frac{3}{2},\frac{{ - {\rho ^2}}}{{4b}}} \right)} \right]\\
+ & \frac{{5{a^3}}}{{432}}{b^{\frac{-7}{6}}}\Gamma \left( {\frac{{ - 5}}{6}} \right)\left[ {1 - {}_2{F_2}\left( {\frac{7}{6},\frac{1}{2};1,\frac{3}{2},\frac{{ - {\rho ^2}}}{{4b}}} \right)} \right]\\
  - & \frac{1}{2}a{b^{\frac{1}{6}}}\Gamma \left( { - \frac{1}{6}} \right)\left[ {1 - {}_2{F_2}\left( {\frac{{ - 1}}{6},\frac{1}{2};1,\frac{3}{2},\frac{{ - {\rho ^2}}}{{4R}}} \right)} \right]\\
 - & \frac{{5{a^3}}}{{5184}}a{b^{\frac{-11}{6}}}\Gamma \left( {\frac{{ - 1}}{6}} \right)\left[ {1 - {}_2{F_2}\left( {\frac{{11}}{6},\frac{1}{2};1,\frac{3}{2},\frac{{ - {\rho ^2}}}{{4b}}} \right)} \right]\\
 + & \frac{1}{4}{a^2}{b^{\frac{{ - 1}}{2}}}\Gamma \left( {\frac{1}{2}} \right)\left[ {1 - {}_2{F_2}\left( {\frac{1}{2},\frac{1}{2};1,\frac{3}{2},\frac{{ - {\rho ^2}}}{{4b}}} \right)} \right]\\
 - & \frac{{{a^3}}}{{320}}{a^2}{b^{\frac{{ - 13}}{2}}}\Gamma \left( {\frac{1}{2}} \right)\left[ {1 - {}_2{F_2}\left( {\frac{5}{2},\frac{1}{2};1,\frac{3}{2},\frac{{ - {\rho ^2}}}{{4b}}} \right)} \right]
\end{split}
\end{equation}
where ${}_p{F_q}\left( {{a_1}, \cdots ,{a_p};{c_1}, \cdots ,{c_q};x} \right)$ is the generalized hypergeometric function, with $p$ and $q$ being positive integers and $\Gamma(\cdot)$ is Gamma function. Noting ${\mathop{\rm Re}\nolimits} \left( { - {\rho ^2}/4b} \right) \gg 1$ and based on the asymptotic behavior of hypergeometric function, i.e., Eq. (8) of \cite{p1_38}, we can approximate \eqref{eq:10} as

\begin{equation}\label{eq:11}
\begin{split}
& \int\limits_0^1 {\sum\limits_{n = 1}^\infty  {\frac{{{{\left( { - 1} \right)}^{n - 1}}{{\left( {\zeta \rho } \right)}^{2n}}}}{{{{\left( {n!} \right)}^2}{2^{2n}}}}\int\limits_0^\infty  {{\kappa ^{2n - \frac{8}{3}}}\exp \left( { - a{\kappa ^{\frac{4}{3}}} - b{\kappa ^2}} \right)d\kappa } } d\zeta}  \\
& \approx  - \frac{1}{2}{b^{\frac{5}{6}}}\Gamma \left( {\frac{{ - 5}}{6}} \right)\frac{{\Gamma \left( {\frac{3}{2}} \right)\Gamma \left( {\frac{4}{3}} \right)}}{{\Gamma \left( {\frac{1}{2}} \right)\Gamma \left( {\frac{{11}}{6}} \right)\Gamma \left( {\frac{7}{3}} \right)}}{\left( {\frac{{{\rho ^2}}}{{4b}}} \right)^{\frac{5}{6}}} = 0.4194{\rho ^{\frac{5}{3}}}
\end{split}
\end{equation}
Similarly, it can be shown that the second and third integrals yield $0.4194{\rho ^{\frac{5}{3}}}$.

Now, we consider the last three integrals which have also identical forms. For the convenience of presentation, consider only the fourth integral. Using Eq. (5) of \cite{p1_36}, this can be expressed as 
\begin{equation}\label{eq:12}
\begin{split}
 \int\limits_0^1 & {\sum\limits_{n = 1}^\infty  {\frac{{{{\left( { - 1} \right)}^{n - 1}}{{\left( {\zeta\rho } \right)}^{2n}}}}{{{{\left( {n!} \right)}^2}{2^{2n}}}}\int\limits_0^\infty  {{\kappa ^{2n - 2}}\exp \left( { - a{\kappa ^{\frac{4}{3}}} - b{\kappa ^2}} \right)d\kappa } } d\zeta} \\
 = & \frac{1}{2}{b^{\frac{1}{2}}}\Gamma \left( { - \frac{1}{2}} \right)\left[ {1 - {}_2{F_2}\left( {\frac{{ - 1}}{2},\frac{1}{2};1,\frac{3}{2},\frac{{ - {\rho ^2}}}{{4b}}} \right)} \right]\\
 + &\frac{{{a^3}}}{{48}}{b^{\frac{-3}{2}}}\Gamma \left( { - \frac{1}{2}} \right)\left[ {1 - {}_2{F_2}\left( {\frac{3}{2},\frac{1}{2};1,\frac{3}{2},\frac{{ - {\rho ^2}}}{{4b}}} \right)} \right]\\
 -& \frac{1}{2}a{b^{ - \frac{1}{6}}}\Gamma \left( {\frac{1}{6}} \right)\left[ {1 - {}_2{F_2}\left( {\frac{1}{6},\frac{1}{2};1,\frac{3}{2},\frac{{ - {\rho ^2}}}{{4b}}} \right)} \right]\\
+& \frac{{7{a^3}}}{{1728}}a{b^{ - \frac{13}{6}}}\Gamma \left( {\frac{1}{6}} \right)\left[ {1 - {}_2{F_2}\left( {\frac{{13}}{6},\frac{1}{2};1,\frac{3}{2},\frac{{ - {\rho ^2}}}{{4b}}} \right)} \right]\\
+ & \frac{1}{4}{a^2}{b^{\frac{{ - 5}}{6}}}\Gamma \left( {\frac{5}{6}} \right)\left[ {1 - {}_2{F_2}\left( {\frac{5}{6},\frac{1}{2};1,\frac{3}{2},\frac{{ - {\rho ^2}}}{{4b}}} \right)} \right]\\
 - & \frac{{11{a^3}}}{{1728}}{a^2}{b^{\frac{{ - 17}}{6}}}\Gamma \left( {\frac{5}{6}} \right)\left[ {1 - {}_2{F_2}\left( {\frac{{17}}{6},\frac{1}{2};1,\frac{3}{2},\frac{{ - {\rho ^2}}}{{4b}}} \right)} \right]
\end{split}
\end{equation}
Noting  ${\mathop{\rm Re}\nolimits} \left( { - {\rho ^2}/4b} \right) \gg 1$ and using the asymptotic behavior of hypergeometric function i.e., Eq. (8) of \cite{p1_38}, we have

\begin{equation}\label{eq:13}
\begin{split}
& \int\limits_0^1 {\sum\limits_{n = 1}^\infty  {\frac{{{{\left( { - 1} \right)}^{n - 1}}{{\left( {\zeta \rho } \right)}^{2n}}}}{{{{\left( {n!} \right)}^2}{2^{2n}}}}\int\limits_0^\infty  {{\kappa ^{2n - 2}}\exp \left( { - a{\kappa ^{\frac{4}{3}}} - b{\kappa ^2}} \right)d\kappa } } d\zeta} \\
& \approx  - \frac{1}{2}{b^{\frac{1}{2}}}\frac{{\Gamma \left( { - \frac{1}{2}} \right)}}{{\Gamma \left( {\frac{1}{2}} \right)}}{\left( {\frac{{{\rho ^2}}}{{4b}}} \right)^{\frac{1}{2}}} = 0.5\rho 
\end{split}
\end{equation}

Replacing  $0.4194{\rho ^{{\raise0.7ex\hbox{$5$} \!\mathord{\left/
 {\vphantom {5 3}}\right.\kern-\nulldelimiterspace}
\!\lower0.7ex\hbox{$3$}}}}$ as the solution of first three integrals and  $0.5\rho$ as the solution of last three integrals, we obtain the final form of wave structure function as
\begin{equation}\label{eq:14}
\begin{split}
W\left( {\rho ,L} \right)  = & 1.44\pi {k^2}L\left( {\frac{{\alpha _{th}^2{\chi _T}}}{{{\omega ^2}}}} \right){\varepsilon ^{ - \frac{1}{3}}}\left( {1.175\eta _K^{2/3}\rho  + 0.419{\rho ^{\frac{5}{3}}}} \right) \\
& \times \left( {{\omega ^2} + {d_r} - \omega \left( {{d_r} + 1} \right)} \right)
\end{split}
\end{equation}
This can be now replaced in \eqref{eq:6} to calculate the average power transfer over the underwater quantum link which is required for the calculation of QBER bound in \eqref{eq:4}.

As a benchmark, we further consider a QKD system operating over the non-turbulent condition. The exact QBER of such a QKD system is given by \cite{p1_31}
\begin{equation}\label{eq:15}
{\rm{QBE}}{{\rm{R}}_{non}} = \frac{{2{n_N}}}{{{n_S}{\mu _0}l + 4{n_N}}}
\end{equation}
where ${\mu _0}$ is the largest eigenvalue of the singular value decomposition of vacuum-propagation Green's function given in \cite{p1_39}.

As a sanity check, we consider two special cases. First, we assume large values of extinction coefficient $\varsigma$ where turbidity effects are more pronounced. For sufficiently large $\varsigma$  values, \eqref{eq:5} can be simplified as $l \approx 0$  and consequently we have $\exp \left( { - \eta {n_S}l} \right) \cong 1$. Furthermore, we assume $\mu  \cong {\mu _0}$ which can be justified for sufficiently short distances where the effect of turbulence is negligible. Replacing these within the derived upper bound on QBER, we obtain
\begin{equation}\label{eq:16}
{\rm{QBER}}\,\, \cong \frac{{2{n_N}}}{{\mu {n_S}l + 4{n_N}}}
\end{equation}
which coincides with \eqref{eq:15} for the non-turbulent case. This result shows that the effect of turbulence is negligible when turbidity is dominant. As a second case, we assume $\mu  \cong {\mu _0} = 1$ which can be justified for short distances in the presence of very weak turbulence. Replacing it within \eqref{eq:4}, we again revert to the non-turbulent case. In other words, when the transmitted photons experience weak turbulence at short distances, the influence of path loss is dominant as expected.

\subsection[SKR Analysis]{SKR Analysis}
SKR is defined as the difference between the amount of information shared by Alice and Bob and the amount of residual information that Eve might have {\cite{p1_30}}. For BB84 protocol, the quantum channel can be modeled as a binary symmetric channel (BSC) with crossover probability of $\rm{QBER}$. The minimum amount of information that should be sent from Alice to Bob in order to correct his key string can be described by the entropy function  $h\left( {{\rm{QBER}}}\right)\!=\!-{\rm{QBER}}{\log _2}\!\left( {{\rm{QBER}}}\right)\!-\left({1-{\rm{QBER}}}\right)\!{\log _2}\left({1-{\rm{QBER}}}\right)$ {\cite{p1_40}}. The amount of disclosed information to Eve in this process can be then expressed as $1 - h\left( {{\rm{QBER}}} \right)$ {\cite{p1_40}}. Therefore, SKR for BB84 protocol can be written as {\cite{p1_40}}
\begin{equation}\label{eq:17}
R = 1 - \left( {1 + f} \right)h\left( {{\rm{QBER}}} \right)
\end{equation}
where $f$ is the reconciliation efficiency {\cite{p1_40}} and its value depends on the employed error correction code.

In our work, we consider low density parity check (LDPC) codes optimized for BSCs {\cite{p1_40}}. The corresponding reconciliation efficiency is given by {\cite{p1_41}}
\begin{equation}\label{eq:18}
f = \frac{{1 - {R_c}}}{{h\left( {{\rm{QBE}}{{\rm{R}}_{th}}} \right)}}
\end{equation}
where ${R_c}$ is the code rate and ${\rm{QBE}}{{\rm{R}}_{th}}$ is a threshold value preset in LDPC code design {\cite{p1_42}}, i.e., this corresponds to the maximum value of  ${\rm{QBER}}$ that can be corrected as the code length tends to infinity. The code rates and threshold QBER values for optimized LDPC codes can be found in Table 1 of {\cite{p1_40}}.

Replacing \eqref{eq:18} in \eqref{eq:17} and using the upper bound on QBER in \eqref{eq:4}, a lower bound on SKR is obtained as
\begin{equation}\label{eq:19}
\begin{aligned}
R \ge & 1 - \left( {1 + \frac{{1 - {R_c}}}{{h\left( {{\rm{QBE}}{{\rm{R}}_{th}}} \right)}}} \right)\\
{\rm{   }} \times {\rm{ }}& h\left(\frac{{{n_N}\left( {1 - \mu  + \mu {e^{ - \eta {n_S}l}}} \right)}}{{\frac{{{n_S}l}}{2}\mu {e^{ - \eta {n_S}l}} + 2{n_N}\left( {1 - \mu  + \mu {e^{ - \eta {n_S}l}}} \right)}} \right)
\end{aligned}
\end{equation}

\section{Numerical Results}\label{sec:P1_Results}
In this section, we demonstrate the performance of underwater QKD scheme under consideration. We assume transmitter beam divergence angle of $\theta=6^{\circ}$, dark current count rate of  $I_{dc} = 60$   $\rm{Hz}$, filter spectral width of $\Delta \lambda = 0.12\times10^{-9}$ $\rm{nm}$, bit period of  $\Delta t = 35$ $\rm{ns}$, receiver gate time of  $\Delta t\ensuremath{'} = 200$ $\rm{ps}$, and Geiger-Mode APD quantum efficiency of  $\eta = 0.5$. Unless otherwise stated, we assume an average photon number of $n_S = 1$, transmitter and receiver aperture diameters of $d_1 = d_2 = 10$ $\rm{cm}$, FOV of  $\Omega = 180^{\circ}$ and clear atmospheric conditions at night with a full moon. We consider clear ocean, coastal water, and harbor water as water types. As for channel parameters, we assume  $\alpha_{th} = 2.56\times10^{-4}$ $\rm{1/deg}$,  $P_T = 7$, $P_S = 6.86\times 10^2$, $P_{TS} = 13.85$ and  $\upsilon = 1.0576\times 10^{-6}$ $\rm{m^{2}s^{-1}}$ \cite{p1_35}. We consider three representative cases for turbulence strength. Specifically, we assume $\omega = -2.2$, $\chi_T = 2\times10^{-7}$ $\rm{K^{2}s^{-3}}$ and $\varepsilon = 2\times10^{-5}$ $\rm{m^{2}s^{-3}}$ for weak turbulence,  $\omega = -2.2$,  $\chi_T = 10^{-6}$ $\rm{K^{2}s^{-3}}$ and $\varepsilon = 5\times10^{-7}$ $\rm{m^{2}s^{-3}}$ for moderate turbulence and  $\omega = -2.2$,  $\chi_T = 10^{-5}$ $\rm{K^{2}s^{-3}}$ and  $\varepsilon = 10^{-5}$ $\rm{m^{2}s^{-3}}$ for strong oceanic turbulence \cite{p1_43}. For the convenience of the reader, the channel and system parameters are summarized in Table \ref{table1}.

\begin{table}[t]
\captionsetup{justification=centering}
\caption{System and channel parameters}
\label{table1}
\begin{center}
\scalebox{0.9}{
\begin{tabular}{ |l|l|l| } 
 \hline
 \textbf{Parameter} & \textbf{Definition} & \textbf{Numerical Value} \\ \hline
$\Omega$ & \text{Field of view} & $180^{\circ}$ \cite{p1_33} \\ \hline 
$\Delta \lambda$ & \text{Filter spectral width} & $30$ $\rm{nm}$ \\ \hline 
$\lambda$ & \text{Wavelength} & $530$ $\rm{nm}$ \cite{p1_33} \\ \hline 
$\Delta t$ & \text{Bit period} & $35$ $\rm{ns}$ \cite{p1_27} \\ \hline 
$\Delta t\ensuremath{'}$ & \text{Receiver gate time}  & $200$ $\rm{ps}$ \cite{p1_27} \\ \hline 
$d_1$ & \text{Transmitter aperture diameter} & $10$ $\rm{cm}$ \cite{p1_31} \\ \hline 
$d_2$ & \text{Receiver aperture diameter} & $10$ $\rm{cm}$ \cite{p1_31} \\ \hline 
$\eta$ & \text{Quantum efficiency} & $0.5$ \cite{p1_31} \\ \hline 
$I_{dc}$ & \text{Dark current count rate} & $60$ $\rm{Hz}$ \cite{p1_27} \\ \hline 
$K_{\infty}$ & \text{Asymptotic diffuse attenuation coefficient} & $0.08$ $\rm{m^{-1}}$ \cite{p1_29} \\ \hline 
$z_d$ & \text{Depth} & $100$ $\rm{m}$ \cite{p1_27} \\ \hline 
$\theta$ & \text{Transmitter beam divergence angle} & $6^{\circ}$ \cite{p1_33} \\ \hline 
$\varsigma$ & \text{Extinction coefficient} \,\begin{tabular}{l|l}  & \text{Clear water} \\ 
 & \text{Coastal water} \\
 & \text{Turbid harbor} \end{tabular} & 
 \begin{tabular}{l} \!\!\!\!$0.151$ $\rm{m^{-1}}$ \cite{p1_32}\\
\!\!\!\! $0.339$ $\rm{m^{-1}}$ \cite{p1_32} \\
\!\!\!\! $2.195$ $\rm{m^{-1}}$ \cite{p1_32} \end{tabular} \\\hline
$T$ & \text{Correction coefficient} \begin{tabular}{c|c}  & $\theta  = 6^\circ ,\,\,{d_1} = 10\,{\rm{cm}}$ \,\,\,\,\,\,\,\,\\
 & $\theta  = 6^\circ ,\,\,{d_1} = 20\,{\rm{cm}}$\,\,\,\,\,\,\,\, \\
 & $\theta  = 6^\circ ,\,\,{d_1} = 30\,{\rm{cm}}$ \,\,\,\,\,\,\,\,\end{tabular} & 
 \begin{tabular}{l}  \!\!\!\! $0.16$  \cite{p1_33}\\
\!\!\!\! $0.21$  \cite{p1_33} \\
\!\!\!\! $0.26$  \cite{p1_33} \end{tabular} \\\hline
\end{tabular}}
\end{center}
\end{table}

\subsection{Effect of water type}
\begin{figure}
     \centering
     \begin{subfigure}[tb]{0.7\linewidth}
         \centering
         \captionsetup{justification=centering}
         \includegraphics[trim=0.3cm 0.7cm 0.5cm 0.5cm,width=\textwidth]{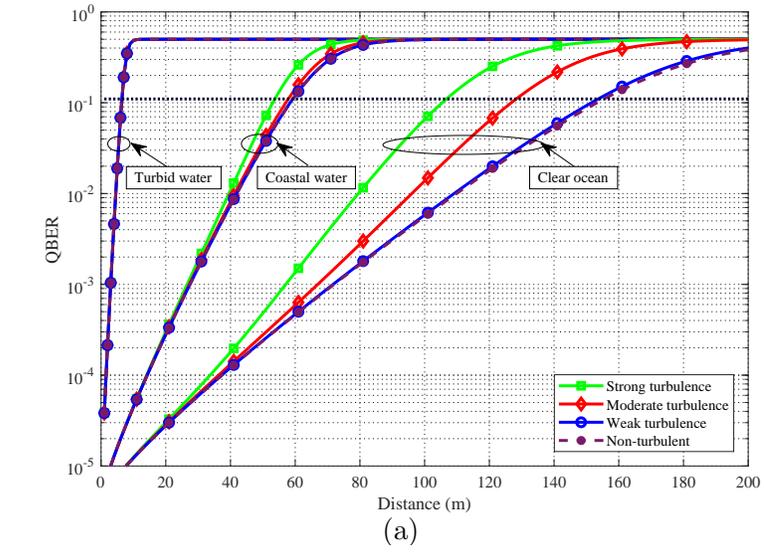}
         \caption{{}}
         \label{fig:fig2_a}
     \end{subfigure}
     \begin{subfigure}[tb]{0.7\linewidth}
         \centering
         \captionsetup{justification=centering}
         \includegraphics[trim=0.3cm 0.65cm 0.5cm 0.1cm,width=\textwidth]{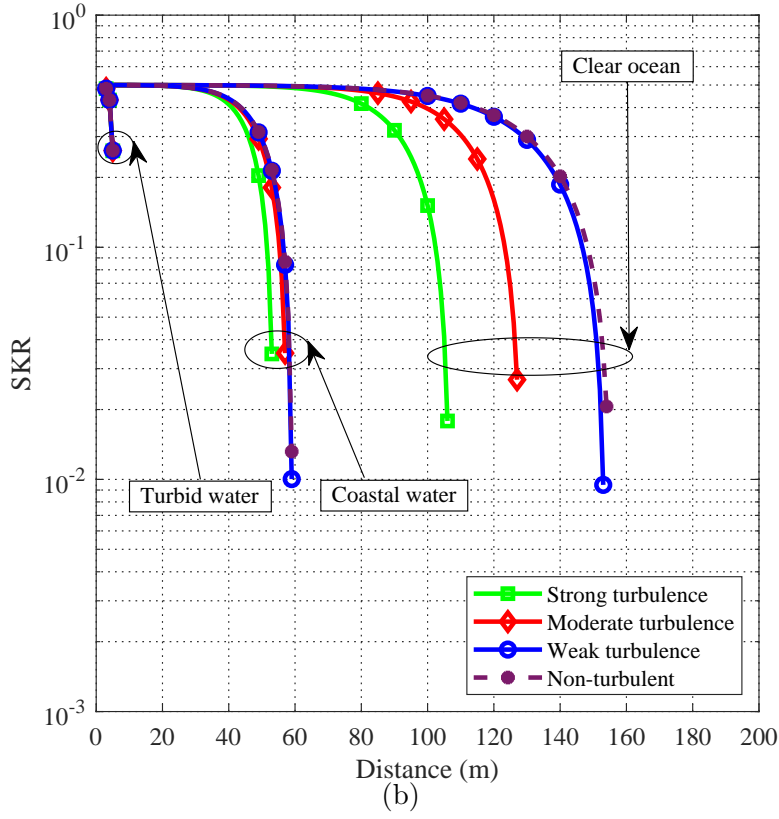}
         \caption{{}}
         \label{fig:fig2_b}
     \end{subfigure}
        \caption{Effect of turbulence strength for different water types at night time with a full moon on (a) QBER (b) SKR}
        \label{fig:2}
\end{figure}

Fig. \ref{fig:2} illustrates the upper bound on QBER and the lower bound on SKR with respect to link distance for clear ocean, coastal water and turbid water. For each water type, we assume weak, moderate and strong turbulence. It can be observed from Fig. \ref{fig:fig2_a} that the turbulence effect in turbid water is negligible and the path loss is the dominant factor which verifies our discussion in Section III. For example, if ${\rm{QBER}} \le {\rm{0.11 }}$ is targeted\footnote{It is generally accepted that for BB84 protocol is secure against a sophisticated quantum attack if QBER is less than 0.11 \cite{p1_44}.}, we have the same achievable distance of $6$ m regardless of the level of turbulence. As turbidity decreases, the achievable distance increases and the effect of turbulence is more pronounced. The QBER performance for non-turbulent case in \eqref{eq:15} is also included as a benchmark. In non-turbulent coastal water, the achievable distance to maintain ${\rm{QBER}} \le {\rm{0.11 }}$  is around $60$ m and reduces to $54$ m for strong turbulence. For clear ocean, the achievable distance for weak turbulence and non-turbulent conditions is the same and around $155$ m confirming our earlier discussion in Section \ref{sec:P1_Per_Analysis}. The achievable distance reduces to $128$ m and $107$ m for moderate and strong turbulence, respectively.


The aforementioned achievable distances are possible under the assumption of perfect error correction. In an effort to have an insight into what transmission distances can be obtained with practical coding schemes, Fig. {\ref{fig:fig2_b}} depicts the SKR performance against the link length. We employ an LDPC code\footnote{In our numerical results, we keep the code rate fixed and employ the LDPC code optimized for QBER threshold value of 0.11. It is also possible to use other LDPC codes in {\cite{p1_40}} optimized for lower QBER values. This will improve SKR, however, the maximum transmission distance will still remain the same because the highest QBER that can be tolerated to obtain non-zero SKR should be less than 0.11.} with a rate of ${R_c} = 0.5$ optimized for a BSC channel with crossover probability of  ${\rm{QBE}}{{\rm{R}}_{th}} = 0.1071 \approx 0.11$ {\cite{p1_40}}. It can be observed that the maximum distance to support a non-zero SKR value for turbid water (regardless of turbulence level) is $5$ m. This is obviously less than the achievable distance of $6$ m obtained through QBER analysis. Similarly, it can be readily checked that the maximum distances to support a non-zero SKR value for other combinations of water type and turbulence level are slightly smaller than achievable distances earlier obtained. For example, the maximum transmission distances for coastal water and clear water in weak turbulence conditions are $59$ m and $153$ m while corresponding QBER analysis yields $60$ m and $155$ m.

\subsection{Effect of atmospheric condition}
\begin{figure}
     \centering
     \begin{subfigure}[b]{0.65\linewidth}
         \centering
         \captionsetup{justification=centering}
         \includegraphics[trim=0.3cm 0.7cm 0.5cm 0.5cm,width=\textwidth]{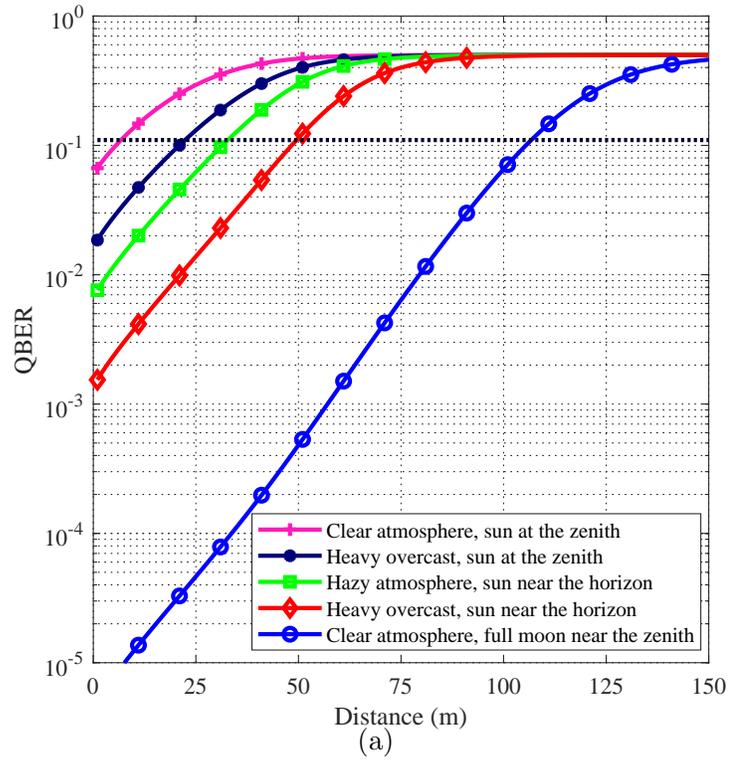}
         \caption{{}}
         \label{fig:fig3_a}
     \end{subfigure}
     \begin{subfigure}[b]{0.65\linewidth}
         \centering
         \captionsetup{justification=centering}
         \includegraphics[trim=0.3cm 0.65cm 0.5cm 0.1cm,width=\textwidth]{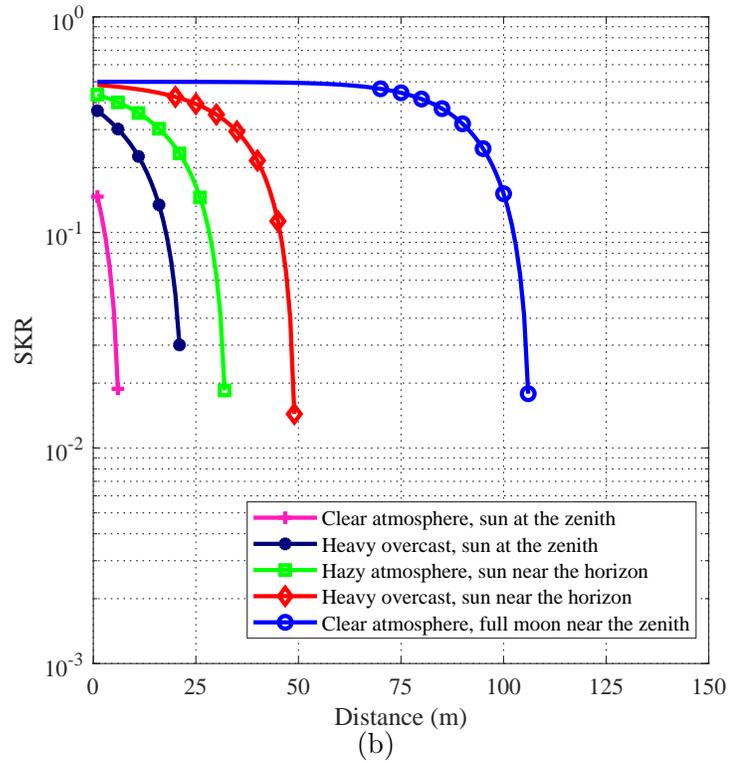}
         \caption{{}}
         \label{fig:fig3_b}
     \end{subfigure}
        \caption{Effect of atmospheric condition in clear ocean with strong turbulence on (a) QBER (b) SKR}
        \label{fig:3}
\end{figure}

In Fig. \ref{fig:3}, we investigate the effect of different atmospheric conditions on the performance of the QKD system. We consider clear ocean with strong turbulence and assume clear, hazy and overcast atmosphere with relative locations of sun and earth at day time. As a benchmark, clear atmospheric conditions at night with a full moon (assumed in Fig. \ref{fig:2}) is also included. It can be observed from Fig. \ref{fig:3} that the achievable distance for underwater QKD system at day time drastically reduces in comparison to night time conditions due to an increase in the received background noise. For example, when the sun is near horizon, the maximum transmission distance (obtained through SKR analysis) for the heavy overcast atmosphere is $49$ m while it reduces to $21$ m and $6$ m respectively for heavy overcast atmosphere and for clear atmosphere with the sun at zenith location. These are much lower than $106$ m achievable under the same system assumptions at night with a full moon.

\subsection{Effect of FOV}
\begin{figure}
     \centering
     \begin{subfigure}[tb]{0.65\linewidth}
         \centering
         \captionsetup{justification=centering}
         \includegraphics[trim=0.3cm 0.7cm 0.5cm 0.5cm,width=\textwidth]{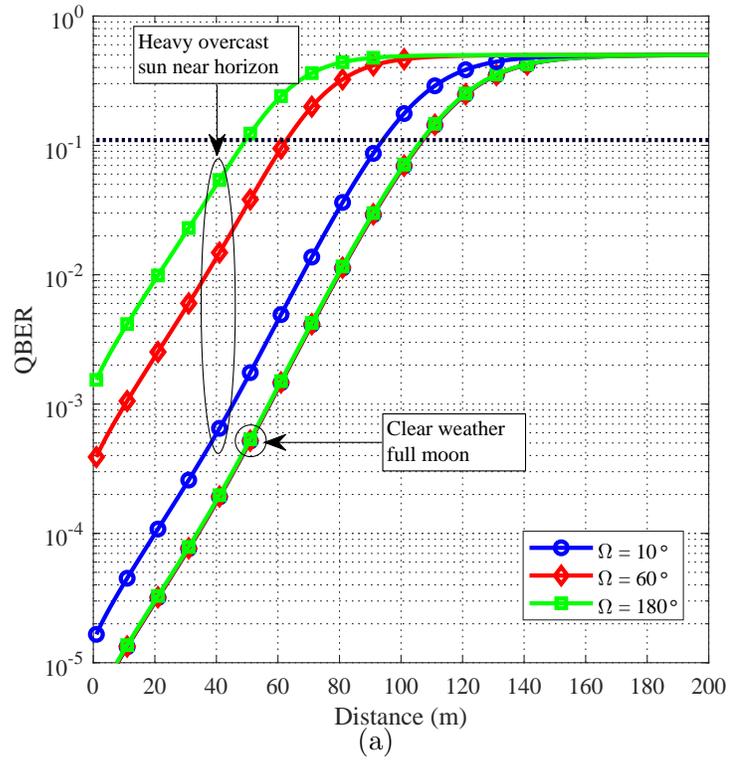}
         \caption{{}}
         \label{fig:fig4_a}
     \end{subfigure}
     \begin{subfigure}[tb]{0.65\linewidth}
         \centering
         \captionsetup{justification=centering}
         \includegraphics[trim=0.3cm 0.65cm 0.5cm 0.1cm,width=\textwidth]{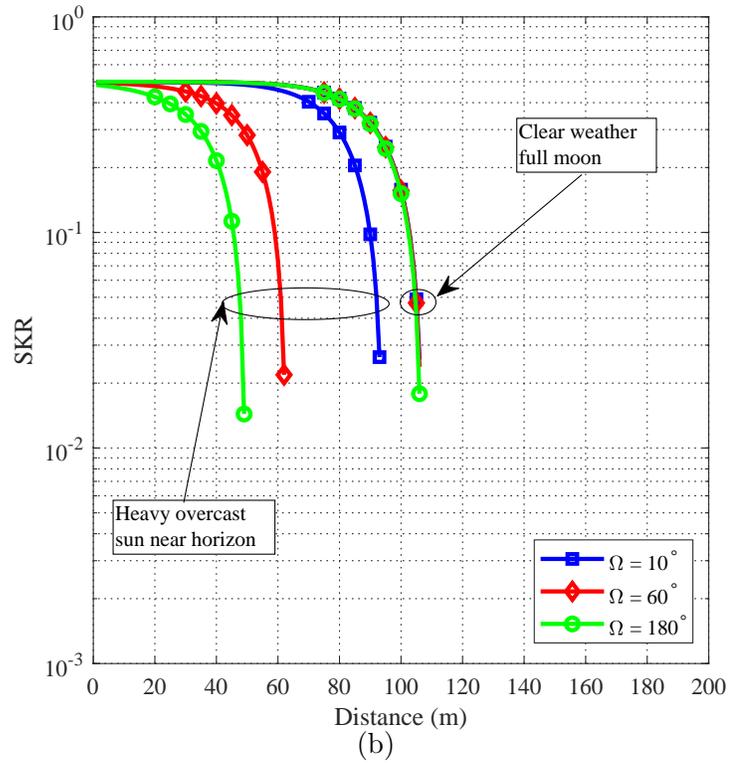}
         \caption{{}}
         \label{fig:fig4_b}
     \end{subfigure}
         \caption{Effect of FOV in clear ocean with strong turbulence under different  atmospheric conditions on (a) QBER (b) SKR}
        \label{fig:4}
\end{figure}

In Fig. \ref{fig:4}, we investigate the effect of FOV on the performance of the QKD system. We assume clear ocean with strong turbulence and consider two atmospheric cases. These are clear weather night with a full moon and heavy overcast when sun is near the horizon. We assume $\Omega = 10^{\circ}$, $60^{\circ}$, and $180^{\circ}$. It is observed that at night time, the effect of FOV is practically negligible and the QBER remains the same for all FOV values under consideration. Benefit of choosing a proper value of FOV becomes clear as the environment irradiance increases. In daylight, we observe that the achievable distance significantly improves as the FOV decreases. This improvement is due to the decrease in background noise as the FOV decreases. Mathematically speaking, the maximum transmission distance (obtained through SKR analysis) for $\Omega = 180^{\circ}$ is around $49$ m, while it increases to $62$ m and $93$ m for $\Omega = 60^{\circ}$ and  $10^{\circ}$, respectively. 

\subsection{Effect of aperture size}
\begin{figure}
     \centering
     \begin{subfigure}[!htb]{0.65\linewidth}
         \centering
         \captionsetup{justification=centering}
         \includegraphics[trim=0.3cm 0.7cm 0.5cm 0.5cm,width=\textwidth]{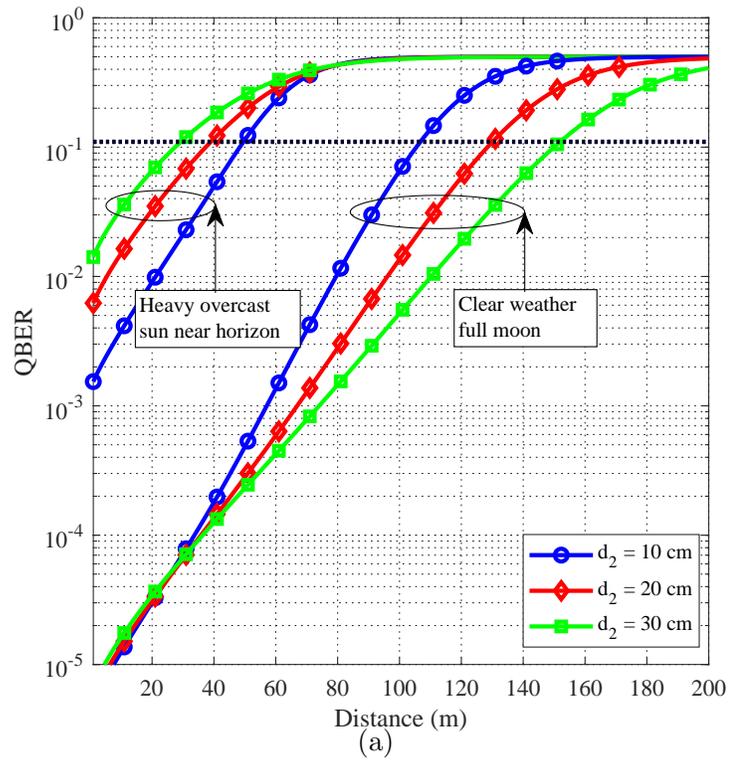}
         \caption{{}}
         \label{fig:fig5_a}
     \end{subfigure}
     \begin{subfigure}[!htb]{0.65\linewidth}
         \centering
         \captionsetup{justification=centering}
         \includegraphics[trim=0.3cm 0.65cm 0.5cm 0.1cm,width=\textwidth]{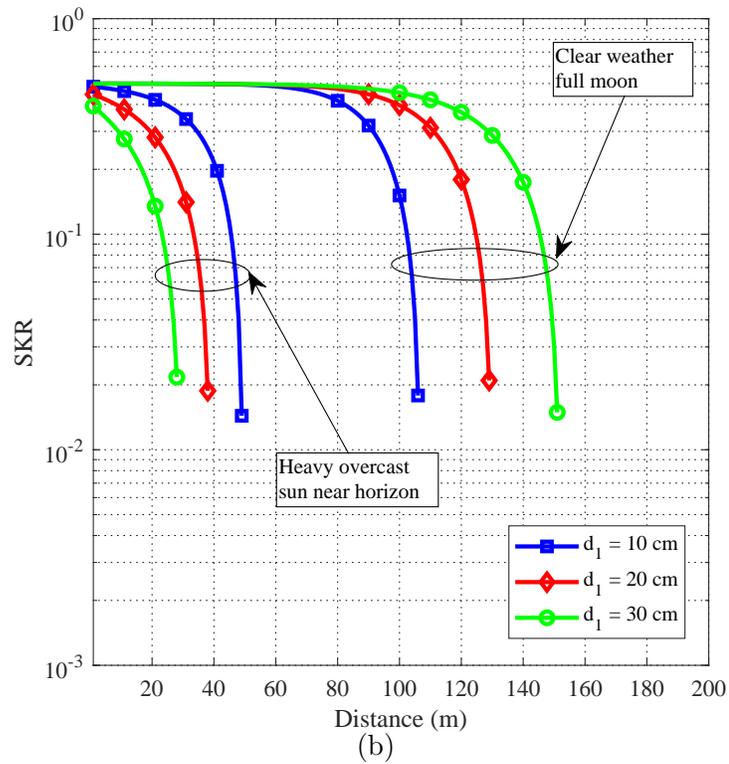}
         \caption{{}}
         \label{fig:fig5_b}
     \end{subfigure}
        \caption{Effect of aperture size in clear ocean with strong turbulence under different atmospheric conditions on  (a) QBER (b) SKR}
        \label{fig:5}
        \vspace*{-0.5cm}
\end{figure}

In Fig. \ref{fig:5}, we study the effect of aperture size on the performance of underwater QKD system. Similar to Fig. \ref{fig:4}, we assume clear ocean with strong turbulence and consider two distinct atmospheric conditions. We assume the receiver aperture size varies as $d_2 = 10$, $20$ and $30$ cm and the transmitter pupil has the same diameter as the receiver. It is observed that at night time with a full moon, the achievable distance increases as the diameter size increases. For example, the maximum transmission distance (obtained through SKR analysis) for $d_2 = 10$ cm is around $106$ m, while it climbs up to $128$ m and $151$ m for $d_2 = 20$ cm and $30$ cm, respectively. It should be emphasized that the increase in background noise as a result of increasing the diameter size is negligible at night. On the other hand, at daylight, increasing the diameter size has a negative effect on the performance. It is observed that the maximum transmission distance for $d_2 = 10$ cm is $49$ m, whereas it decreases to $38$ m and $27$ m for diameter size of $20$ and $30$ cm, respectively. These observations indicate the necessity of using adaptive selection of aperture size in practical implementations.

\chapter{Multi-Hop QKD with Passive Relays over Underwater Turbulence Channels}\label{ch:relay}

In this chapter, we consider a multi-hop underwater QKD system where relay nodes are utilized along the path connecting two legitimate parties. Unlike classical optical communication systems \cite{p2_18}, amplify-and-forward and detect-and-forward relaying cannot be used in QKD since any type of measurement modifies the quantum state \cite{p2_3}. To address this, we utilize passive relays \cite{p2_17} which simply redirect the qubits to the next relay node or to the destination node without performing any measurement or detection process. Under the assumption of passive relays and based on a near-field analysis \cite{p1_31} over underwater turbulence channels, we derive an upper bound on QBER and a lower bound on SKR. Based on these bounds, we present the performance of underwater QKD in different water types and different levels of turbulence strength. We further investigate the effect of system parameters such as detector field-of-view (FOV) and aperture size on the system performance. Our results demonstrate that the multi-hop schemes with judiciously selected values of relay number, FOV size and aperture size successfully improve the end-to-end distance in water types with low turbidity.

The remainder of this chapter is organized as follows. In Section \ref{sec:P2_sys}, we describe our relay-assisted system model based on BB84 QKD protocol. In Section \ref{sec:P2_Performance_Analysis}, we derive an upper bound on the QBER and a lower bound on the SKR of the system. In Section \ref{sec:P2_Results}, we present numerical results.

\section{System Model}\label{sec:P2_sys}
\begin{figure}[t]
\centering
\includegraphics[width=0.8\textwidth]{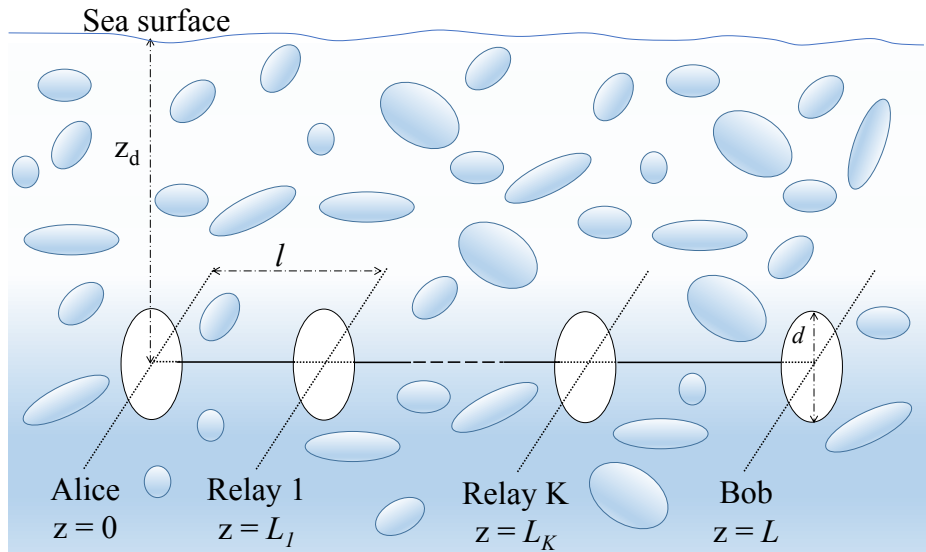}
\caption{Schematic diagram of the underwater relay-assisted QKD system with $K$ relay nodes.}
\label{fig:p_2_1}
\end{figure}

We consider a relay-assisted underwater QKD system with $K$ serial passive relay nodes over a link distance of $L$. As illustrated in Fig. \ref{fig:p_2_1}, Alice (transmitter) with a diameter size of $d$ is placed in $z = 0$ plane. Relay nodes and Bob (receiver) with the same diameter size of $d$ are located in the $z = {L_i}$. The consecutive nodes in the serial scheme are placed equidistant along the path from the source to the destination. Therefore, the length of each hop is equal to $l = {L \mathord{\left/
 {\vphantom {L {\left( {K + 1} \right)}}} \right.
 \kern-\nulldelimiterspace} {\left( {K + 1} \right)}}$.

The QKD system is built upon BB84 protocol \cite{p1_8} which aims to create a secret key between the authorized partners (Alice and Bob) such that eavesdropper (Eve) fails to acquire meaningful information. BB84 protocol is based on the principle of polarization encoding. In this protocol, Alice prepares a qubit by choosing randomly between two linear polarization bases namely rectilinear (denoted by $ \oplus $) or diagonal (denoted by $ \otimes $) for every bit she wants to send. She selects a random bit value \qq{0} or \qq{1} and uses polarization encoding of photons where polarization of ${{0^\circ } \mathord{\left/
 {\vphantom {{0^\circ } { - 45^\circ }}} \right.
 \kern-\nulldelimiterspace} { - 45^\circ }}$ represents 0 and polarization of ${{ + 90^\circ } \mathord{\left/
 {\vphantom {{ + 90^\circ } { + 45^\circ }}} \right.
 \kern-\nulldelimiterspace} { + 45^\circ }}$ represents 1. At the receiver side, Bob measures the arriving photon randomly in either $ \oplus $ or $ \otimes $ bases. Alice and Bob determine the secure key with respect to the received qubits at the \qq{sift} events. Sift events occurs at the bit intervals in which exactly one of the single photon detectors registers a count and both Alice and Bob have chosen the same basis. Alice and Bob can recognize the sift events by transferring information over a public classical communication channel (underwater optical channel in our case). Based on the sifted qubits, a shared one-time pad key is created to use for secure communication \cite{p1_26}.

Alice generates each qubit with an average photon number of ${n_S}$ which is encoded with the corresponding polarization state of the qubit for a randomly chosen basis. As a result of underwater path loss and turbulence, the ${i^{th}}$ relay ($i = 1, \ldots ,K$) collects only a random fraction ${\gamma _i}$ of the transmitted photons. Under the assumption of identical transmitter/receiver sizes and equidistant placement, we can simply write ${\gamma _1} = {\gamma _2} = .... = {\gamma _K} = \gamma $. The relay node forwards the captured photons to the next relay (or Bob) by redirecting the light beam and without any measurements. Therefore, Bob will collect an overall fraction ${\gamma _{Bob}} = {\gamma ^{K + 1}}$ of the originally transmitted photons from Alice.

In addition to the received photons from the source, receiver of each relay node will collect some background noise. The total average number of background photons per polarization at the ${i^{th}}$ relay can be therefore expressed as
\begin{equation}\label{eq:p2_1}
{n_{{B_i}}} = {n_{B0}} + {n_{B0}}\gamma  + {n_{B0}}{\gamma ^2} +  \ldots  + {n_{B0}}{\gamma ^{i - 1}}
\end{equation}

The accumulated background photons at Bob's receiver becomes \cite{p2_17}
\begin{equation}\label{eq:p2_2}
{n_B} = {n_{B0}}\frac{{1 - {\gamma ^{K + 1}}}}{{1 - \gamma }}    
\end{equation}

Beside background noise, each of Bob's detectors will collect dark current noise. Let ${I_{dc}}$ and $\Delta t$ denote the dark current count rate and the bit period, respectively. The average number of dark counts is given by ${n_D} = {I_{dc}}\Delta t$. Thus, the average number of noise photons reaching each Bob's detector can be obtained by ${n_N} = {{{n_B}} \mathord{\left/
 {\vphantom {{{n_B}} 2}} \right.
 \kern-\nulldelimiterspace} 2} + {n_D}$. It should be noted that since the relays just redirect the photons, they do not increase the dark current.

\section{Performance Analysis}\label{sec:P2_Performance_Analysis}
In this section, we investigate the performance of the underwater QKD system through the derivation of an upper bound on QBER and a lower bound on the SKR.

\subsection{QBER Analysis}
QBER is the ratio of probabilities of sift and error which depend on the statistical characteristics of 
the capture fraction $\gamma $ (i.e., received fraction of transmitted photons). The capture fraction can be obtained based on the extended Huygens-Fresnel principle {\cite{p1_31}}. As discussed in {\cite{p1_31}}, in order to calculate the received field, we need to determine the paraxial Green's function which is a function of the complex phase perturbation of the field describing the turbulence of the path. To make the analysis mathematically tractable, Green's function can be replaced with an equivalent set of fictitious parallel channels by normal mode decomposition. Let  $\hat \mu $ denote the largest eigenvalue.
This is also called as \qq{power transfer} in \cite{p2_17, p2_22} and defines the probability of transmitted photon being reliably received in the presence of turbulence.
The statistical description of  $\hat \mu $ is unfortunately not available in the literature.
As an alternative, an upper bound on QBER was presented in \cite{p2_17} using an upper bound on the noise count and a lower bound on the maximum average power transfer, i.e., $\mu  \buildrel \Delta \over = {\mathop{\rm E}\nolimits} \left[ {\hat \mu } \right]$.
Specifically, this is given by

\begin{equation}
 {\rm{QBER}} \le \frac{{2\eta {{\hat n}_N}\exp \left[ { - \eta 4{{\hat n}_N}} \right]\left( {1 - {\mu ^{K + 1}} + \exp \left[ { - \eta {n_S}{h^{K + 1}}\left( l \right)} \right]{\mu ^{K + 1}}} \right)}}{{b\exp \left[ { - b} \right]\left( {1 - c} \right) + \left( {a + b} \right)\exp \left[ { - \left( {a + b} \right)} \right]c}}
 \label{eq:p2_3}
\end{equation}

In \eqref{eq:p2_3}, $\eta $ is the quantum efficiency of Geiger-mode avalanche photodiodes (APDs), $\mu  \buildrel \Delta \over = {\mathop{\rm E}\nolimits} \left[ {\hat \mu } \right]$ is the average power transfer and ${\hat n_N}$ is an upper bound on the noise count, i.e., ${n_N} \le {\hat n_N}$, whose derivation will be elaborated later. In \eqref{eq:p2_3}, $a$, $b$ and $c$ are respectively defined by
\begin{equation}\label{eq:p2_4}
a = \eta \left[ {{n_S}{h^{K + 1}}\left( l \right) + 2{n_{B0}}\left( {\frac{{1 - {h^{K + 1}}\left( l \right)}}{{1 - h\left( l \right)}} - 1} \right)} \right]    
\end{equation}
\begin{equation}\label{eq:p2_5}
b = \eta \left( {2{n_{B0}} + 4{n_D}} \right)    
\end{equation}
\begin{equation}\label{eq:p2_6}
c = \frac{{{n_S}{{\left( {\mu h\left( l \right)} \right)}^{K + 1}} + 2{n_{B0}}\left( {\frac{{1 - {{\left( {\mu h\left( l \right)} \right)}^{K + 1}}}}{{1 - \mu h\left( l \right)}} - 1} \right)}}{{{n_S}{h^{K + 1}}\left( l \right) + 2{n_{B0}}\left( {\frac{{1 - {h^{K + 1}}\left( l \right)}}{{1 - h\left( l \right)}} - 1} \right)}}    
\end{equation}

The calculation of $h\left( l \right)$, $\mu $ and ${\hat n_N}$ depends on the operation environment and therefore earlier results in the literature reported for other propagation environments \cite{p2_17} are not applicable. In the following, we elaborate on their calculations for the underwater channel under consideration.

\textbf{Underwater path loss: } For collimated light sources, the geometrical loss is negligible; therefore, the path loss for a laser diode transmitter only depends on the attenuation loss including the effects of absorption and scattering. Based on the modified version of Beer-Lambert formula \cite{p1_33}, the underwater path loss can be expressed as

\begin{equation}\label{eq:p2_7}
h\left( l \right) = \exp \left[ { - \varsigma l{{\left( {\frac{d}{{\theta l}}} \right)}^T}} \right]   
\end{equation}
where $\varsigma $ is extinction coefficient, $l$ is transmission distance, $\theta $ is the full-width transmitter beam divergence angle and $T$ is a correction coefficient based on water type \cite{p1_33}. Extinction coefficient depends on the wavelength and water type. Typical values of extinction coefficients for $\lambda = 532$ nm (green color) in different water types can be found in \cite{p1_32}.

\textbf{Underwater average power transfer: }
The average power transfer for each hop (i.e., a distance of $l$ over turbulent path) can be expressed as \cite{p1_31}
 \begin{equation}\label{eq:p2_8}
     \mu \! =\!\!\frac{{8\sqrt F }}{\pi }\!\!\int\limits_0^1 \!{e^{(\frac{- W({dx,l})}{2})}
     \!\left( {{{\cos }^{ - 1}}\left(\! x \right)\! -\! x\sqrt {1 - {x^2}} }\! \right)\!{J_1}\left( {4x\sqrt F } \right)\!dx}
 \end{equation}
where ${J_1}\left( \boldsymbol{\cdot}   \right)$ is the first-order Bessel function of the first kind and $F$ is the Fresnel number product of transmit and receive diameters given by $F = {\left( {\pi {d^2}/4\lambda l} \right)^2}$. In \mbox{\eqref{eq:p2_8}}, $W\left( \boldsymbol{\cdot} , \boldsymbol{\cdot}   \right)$ is the underwater wave structure function. For a given transmission distance of $l$ and given separation distance between two points on the phase front transverse to the axis of propagation (denoted by $\rho $), it is expressed as \cite{p2_14}
\begin{equation}\label{eq:p2_9}
\begin{split}
    W\left( {\rho ,l} \right) = & 1.44\pi {k^2}l\left( {\frac{{{\alpha ^2}\chi }}{{{\omega ^2}}}} \right){\varepsilon ^{ - \frac{1}{3}}}\left( {1.175\eta _K^{2/3}\rho  + 0.419{\rho ^{\frac{5}{3}}}} \right) \\
    &\times \left( {{\omega ^2} + {d_r} - \omega \left( {{d_r} + 1} \right)} \right)
    \end{split}
\end{equation}
where $k = {{2\pi } \mathord{\left/
 {\vphantom {{2\pi } \lambda }} \right.
 \kern-\nulldelimiterspace} \lambda }$ is the wave number, $\omega $ is the relative strength of temperature and salinity fluctuations, $\varepsilon $ is the dissipation rate of turbulent kinetic energy per unit mass of fluid, $\alpha $ is the thermal expansion coefficient, $\chi $ is the dissipation rate of mean-squared temperature and ${d_r}$ is the eddy diffusivity ratio. In \eqref{eq:p2_9}, ${\eta _K}$ is Kolmogorov microscale length and given by ${\eta _K} = {\left( {{{{\upsilon ^3}} \mathord{\left/
 {\vphantom {{{\upsilon ^3}} \varepsilon }} \right.
 \kern-\nulldelimiterspace} \varepsilon }} \right)^{{1 \mathord{\left/
 {\vphantom {1 4}} \right.
 \kern-\nulldelimiterspace} 4}}}$ with $\upsilon $ referring to the kinematic viscosity. 

\textbf{Underwater noise count: }
In an underwater environment, the primary source of noise is the refracted sunlight from the surface of the water. Let ${R_d}\left( {\lambda ,{z_d}} \right)$ denote the irradiance of the underwater environment as a function of wavelength and underwater depth. With respect to sea surface (i.e., ${z_d} =0$), it can be written as ${R_d}\left( {\lambda ,{z_d}} \right) = {R_d}\left( {\lambda ,0} \right){e^{ - {K_\infty }{z_d}}}$ where ${K_\infty }$ is the asymptotic value of the spectral diffuse attenuation coefficient for spectral down-welling plane irradiance \cite{p1_28}. The background photons per polarization on average can be then given by \cite{p1_27}
\begin{equation}\label{eq:p2_10}
{n_{B0}} = \frac{{\pi {R_d}A\Delta t'\lambda \Delta \lambda \left( {1 - \cos \left( \Omega \right)} \right)}}{{2{h_p}{c_0}}}
\end{equation}
where $A$ is the receiver aperture area, $\Omega $ is the FOV of the detector, ${h_p}$ is Planck's constant, ${c_0}$ is the speed of light, $\Delta \lambda $ is the filter spectral width, and $\Delta t'$ is the receiver gate time. Ignoring the effect of turbulence (i.e., $\hat \mu = 1$) on the redirected background photons coming from relays \cite{p2_17}, an upper bound on the noise count at each of Bob's four detectors can be obtained as
\begin{equation}\label{eq:p2_11}
{\hat n_N} = \frac{{{n_{B0}}}}{2}\left( {\frac{{1 - \exp \left[ { - \varsigma {L^{1 - T}}{{\left( {\frac{{d\left( {K + 1} \right)}}{\theta }} \right)}^T}} \right]}}{{1 - \exp \left[ { - \varsigma {{\left( {\frac{L}{{K + 1}}} \right)}^{1 - T}}{{\left( {\frac{d}{\theta }} \right)}^T}} \right]}}} \right) + {n_D}   
\end{equation}
Replacing \eqref{eq:7}, \eqref{eq:p2_8} and \eqref{eq:p2_11} in \eqref{eq:p2_3}, we can obtain the upper bound on QBER for underwater environments.

\textbf{Special case: }
As a sanity check, consider $K=0$ (i.e., no relay). Therefore, ${\hat n_N}$, $a$, and $c$ can be simplified as ${\hat n_N} = {{{n_{B0}}} \mathord{\left/
 {\vphantom {{{n_{B0}}} 2}} \right.
 \kern-\nulldelimiterspace} 2} + {n_D} = {b \mathord{\left/
 {\vphantom {b {4\eta }}} \right.
 \kern-\nulldelimiterspace} {4\eta }}$, $a = \eta {n_S}h\left( L \right)$ and $c = {\mu _L}$. Replacing these in \eqref{eq:p2_3} yields
 \begin{equation}\label{eq:p2_12}
{\rm{QBER}} \le \frac{{{{\hat n}_N}\left[ {1 \!- \!{\mu _L} + \!{\mu _L}{e^{ - \eta {n_S}h\left( L \right)}}} \right]}}{{\frac{{{n_S}{\mu _L}h\left( L \right)}}{2}{e^{ - \eta {n_S}h\left( L \right)}} \!+ \!2{{\hat n}_N}\!\left[ {1 \!- \!{\mu _L} + \!{\mu _L}{e^{ - \eta {n_S}h\left( L \right)}}} \right]}}
 \end{equation}
where $h\left( L \right)$ and ${\mu _L}$ are respectively the path loss and the average power transfer for the length of direct link connecting Alice and Bob. It can be readily checked that this result coincides with [Eq. (4), 14] which was earlier reported for underwater QKD link.

\subsection{SKR Analysis}
SKR is the difference between the amount of information shared by Alice and Bob and the amount of residual information that Eve might have \cite{p1_30}. In SKR analysis, the quantum channel in BB84 protocol can be modeled as a binary symmetric channel where QBER defines the crossover probability. The minimum amount of information that should be sent from Alice to Bob in order to correct his key string can be described by the entropy function $H\left( {{\rm{QBER}}} \right)\! = \!- {\rm{QBER}}{\log _2}\left( {{\rm{QBER}}} \right)\! - \! \left( {1 \!- \!{\rm{QBER}}} \right){\log _2}\left( {1\! -\! {\rm{QBER}}} \right)$. The amount of disclosed information to Eve can be then expressed as $1 - H\left( {{\rm{QBER}}} \right)$ \cite{p1_40}. In practice, the effect of error correction should be further taken into account. Therefore, we can write the SKR as
\begin{equation}\label{eq:p2_13}
R = 1 - H({\rm{QBER}}) - f \times H({\rm{QBER}})
\end{equation}
where $f > 1$ is the reconciliation efficiency of the error correction scheme \cite{p1_40}. 

Here, we adopt low-density parity check (LDPC) codes optimized for BSCs \cite{p1_40} with a reconciliation efficiency of $f = \left( {1 - {R_c}} \right)/H\left( {{\rm{QBE}}{{\rm{R}}_{th}}} \right)$ \cite{p1_41}. Here, ${R_c}$ denotes the code rate and ${\rm{QBE}}{{\rm{R}}_{th}}$ is a threshold value preset in LDPC code design \cite{p1_42}. Replacing the definition of $f$ and the upper bound on QBER given by \eqref{eq:p2_3} in \eqref{eq:p2_13}, we obtain a lower bound for the SKR as

\begin{equation}
 R \ge 1 - \left( {1 + \frac{{1 - {R_c}}}{{H\left( {{\rm{QBE}}{{\rm{R}}_{th}}} \right)}}} \right)H\left( {\frac{{\eta {{\hat n}_N}\exp \left[ { - \eta 4{{\hat n}_N}} \right]\left( {1 - {\mu ^{K + 1}} + {\mu ^{K + 1}}\exp \left[ { - \eta {n_S}{h^{K + 1}}\left( l \right)} \right]} \right)}}{{\frac{b}{2}\exp \left[ { - b} \right]\left( {1 - c} \right) + \left( {\frac{{a + b}}{2}} \right)\exp \left[ { - \left( {a + b} \right)} \right]c}}} \right)
 \label{eq:p2_14}
\end{equation}

\section{Numerical Results}\label{sec:P2_Results}
In this section, we demonstrate the performance of relay-assisted underwater QKD scheme under consideration in terms of QBER and SKR. We assume the transmitter beam divergence angle of $\theta=6^{\circ}$, dark current count rate of $I_{dc} = 60$ $\rm{Hz}$, filter spectral width of $\Delta \lambda = 30$ $\rm{nm}$, bit period of $\Delta t = 35$  $\rm{ns}$, receiver gate time of  $\Delta t\ensuremath{'} = 200$ $\rm{ps}$, average photon number of $n_S = 1$, and Geiger-Mode APD quantum efficiency of $\eta = 0.5$.
Unless otherwise stated, we assume the transmitter and receiver aperture diameters of $d = 5$ $\rm{cm}$, FOV of $\Omega = 180^{\circ}$, a depth of  ${z_d} = $ 100 m and clear atmospheric conditions at night with a full moon. The typical total irradiances at sea level, i.e., ${R_d}\left( {\lambda ,0} \right)$, in the visible wavelength band for some typical atmospheric conditions are provided in \cite{p1_29}. As for channel parameters, we assume ${\alpha = 2.56\times10^{-4} \: \rm{1/deg}}$ and ${\upsilon = 1.0576\times 10^{-6}\: \rm{m^{2}s^{-1}}}$. We consider two representative cases for turbulence strength. Specifically, we assume $\omega = -2.2$, ${\chi_T = 10^{-6} \: \rm{K^{2}s^{-3}}}$ and ${\varepsilon = 5\times10^{-7} \rm{m^{2}s^{-3}}}$ for moderate turbulence and $\omega = -2.2$, ${\chi_T = 10^{-5} \: \rm{K^{2}s^{-3}}}$ and ${\varepsilon = 10^{-5} \: \rm{m^{2}s^{-3}}}$ \cite{p1_43}. For the convenience of the reader, the channel and system parameters are summarized in Table \ref{table_p2_1}.

\begin{table}
\captionsetup{justification=centering}
\caption{System and channel parameters}
\label{table_p2_1}
\begin{center}
\scalebox{0.8}{
\begin{tabular}{ |l|l|l| } 
 \hline
 \textbf{Parameter} & \textbf{Definition} & \textbf{Numerical Value} \\ \hline
$\Omega$ & \text{Field of view} & $180^{\circ}$ \cite{p1_33} \\ \hline 
$\Delta \lambda$ & \text{Filter spectral width} & $30$ $\rm{nm}$ \cite{p2_14} \\ \hline 
$\lambda$ & \text{Wavelength} & $530$ $\rm{nm}$ \cite{p1_33} \\ \hline 
$\Delta t$ & \text{Bit period} & $35$ $\rm{ns}$ \cite{p1_27} \\ \hline 
$\Delta t\ensuremath{'}$ & \text{Receiver gate time}  & $200$ $\rm{ps}$ \cite{p1_27} \\ \hline 
$d$ & \text{Transmitter aperture diameter} & $5$ $\rm{cm}$ \cite{p1_31} \\ \hline 
$d'$ & \text{Receiver aperture diameter} & $5$ $\rm{cm}$ \cite{p1_31} \\ \hline 
$\eta$ & \text{Quantum efficiency} & $0.5$ \cite{p1_31} \\ \hline 
$I_{dc}$ & \text{Dark current count rate} & $60$ $\rm{Hz}$ \cite{p1_27} \\ \hline 
$K_{\infty}$ & \text{Asymptotic diffuse attenuation coefficient} & $0.08$ $\rm{m^{-1}}$ \cite{p1_28} \\ \hline 
$z_d$ & \text{Depth} & $100$ $\rm{m}$ \cite{p1_27} \\ \hline 
$\theta$ & \text{Transmitter beam divergence angle} & $6^{\circ}$ \cite{p1_33} \\ \hline 
$\varsigma$ & \text{Extinction coefficient} \,\begin{tabular}{l|l}  & \text{Clear water} \\ 
 & \text{Coastal water} \end{tabular} & 
 \begin{tabular}{l} \!\!\!\!$0.151$ $\rm{m^{-1}}$ \cite{p1_32}\\
\!\!\!\! $0.339$ $\rm{m^{-1}}$ \cite{p1_32} \end{tabular} \\\hline
$T$ & \text{Correction coefficient} \begin{tabular}{c|c}  & $\theta  = 6^\circ ,\,\,{d_1} = 5\,{\rm{cm}}$ \,\,\,\,\,\,\,\,\\
 & $\theta  = 6^\circ ,\,\,{d_1} = 10\,{\rm{cm}}$ \,\,\,\,\,\,\,\,\\
 & $\theta  = 6^\circ ,\,\,{d_1} = 20\,{\rm{cm}}$\,\,\,\,\,\,\,\, \\
 & $\theta  = 6^\circ ,\,\,{d_1} = 30\,{\rm{cm}}$ \,\,\,\,\,\,\,\,\end{tabular} & 
 \begin{tabular}{l}  \!\!\!\! $0.13$  \cite{p1_33}\\
 \!\!\!\! $0.16$  \cite{p1_33}\\
\!\!\!\! $0.21$  \cite{p1_33} \\
\!\!\!\! $0.26$  \cite{p1_33} \end{tabular} \\\hline
\end{tabular}}
\end{center}
\end{table}

\begin{figure}
     \centering
     \begin{subfigure}{0.7\textwidth}
         \centering
         \captionsetup{justification=centering}
         \includegraphics[trim=0cm 0.3cm 0.1cm 0.6cm, clip, width=\linewidth]{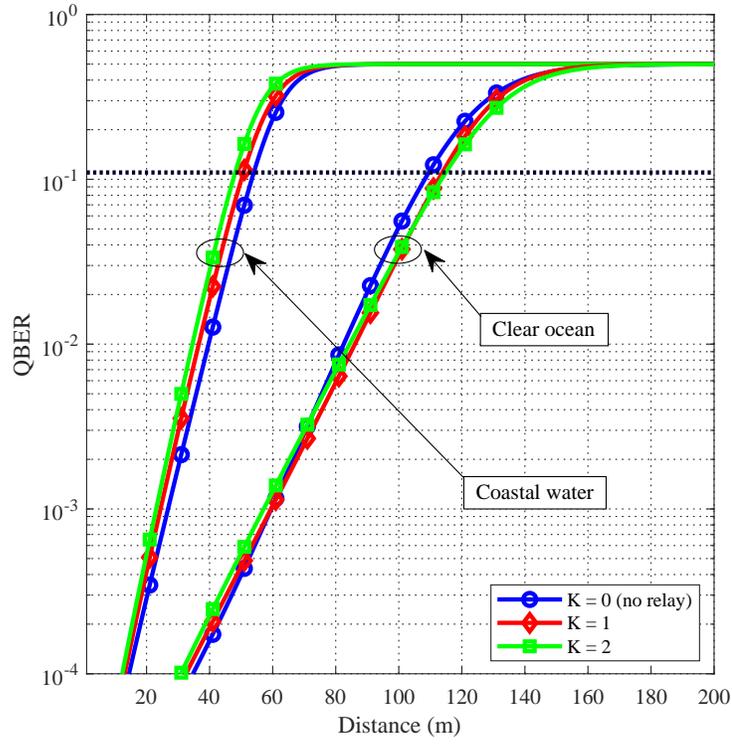}
         \caption{{}}
         \label{fig:fig2_p2_a}
     \end{subfigure}
     \begin{subfigure}{0.7\textwidth}
         \centering
         \captionsetup{justification=centering}
         \includegraphics[trim=0cm 0.3cm 0.1cm 0.6cm, clip, width=\linewidth]{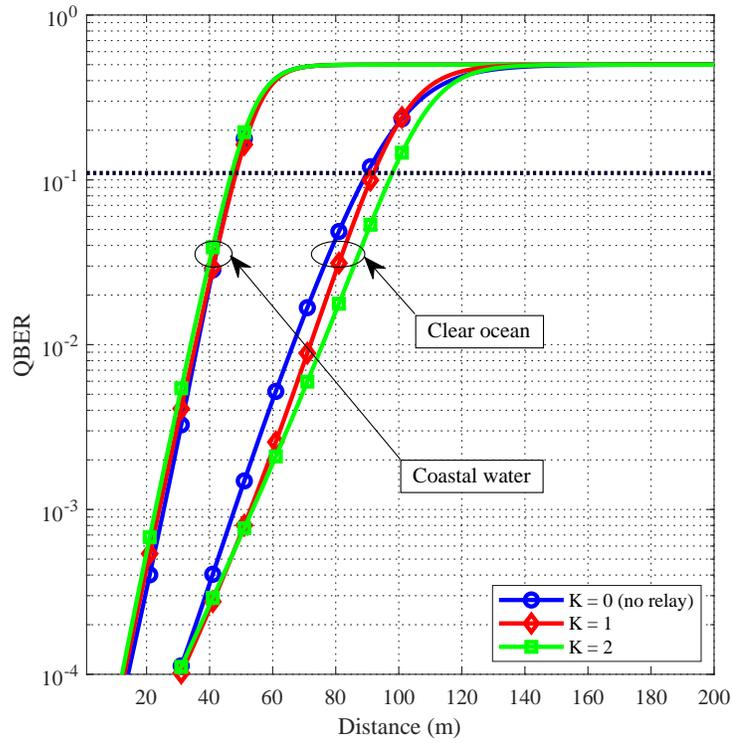}
         \caption{{}}
         \label{fig:fig2_p2_b}
     \end{subfigure}
        \caption{QBER of the relay-assisted QKD system over clear ocean and coastal water for \textbf{(a)} moderate turbulence conditions \textbf{(b)} strong turbulence conditions. }
        \label{fig:p_2_2}
\end{figure}

In Fig. \ref{fig:p_2_2}, we illustrate the performance of QBER with respect to end-to-end link distance assuming different water types (based on turbidity level\footnote{Turbid water results in large extinction coefficient value while the extinction coefficient in non-turbid water takes small values. Based on typical chlorophyll concentrations, pure sea and clear ocean are considered as non-turbid water and the coastal and harbor can be considered as turbid water.}) and turbulence conditions. In our simulations, we consider clear ocean and coastal water whose extinction coefficients are $0.151$  $\rm{m^{-1}}$ and $0.339$ $\rm{m^{-1}}$, respectively. We consider relay-assisted systems with $K = 1$, and $2$ relay nodes. The results for direct link, i.e. $K = 0$, are further included as benchmarks.

It is observed from Fig. \ref{fig:p_2_2} that relaying is not beneficial and even detrimental in turbid water (coastal water). To understand the reasons behind this, it should be noted that there is a fundamental trade-off between accumulated noise and the average number of collected photons coming from Alice. Adding passive relay leads to additional collected background noise redirected from relays to Bob's receiver. Although shorter hops decrease the photon loss caused by turbulence, it is not always able to mitigate the exponential loss of photons due to the path loss. Specifically, in turbid water where the value of the extinction coefficient is large, the turbulence effect on the QBER performance becomes insignificant with respect to the path loss.

On the other hand, relaying helps improve the performance for non-turbid water as seen in the plots associated with clear ocean. For instance, in moderate turbulence conditions (Fig. \ref{fig:fig2_p2_a}) to achieve ${\rm{QBER}} \le 0.11$\footnote{It is generally accepted that for BB84 protocol is secure against a sophisticated quantum attack if QBER is less than 0.11 \cite{p1_44}.}, the achievable distance for direct link is $109$ m. It increases to $113$ m and $114$ m with $K = 1$ and $K = 2$ relay nodes, respectively. The improvements are more pronounced as turbulence strength increases. In strong turbulence conditions (Fig. \ref{fig:fig2_p2_b}) to achieve ${\rm{QBER}} \le 0.11$, the achievable distance for direct link is $89$ m. It increases to $91$ m and $97$ m with $K = 1$ and $K = 2$ relay nodes, respectively.

\begin{figure}[t]
\centering
\includegraphics[width=0.7\textwidth]{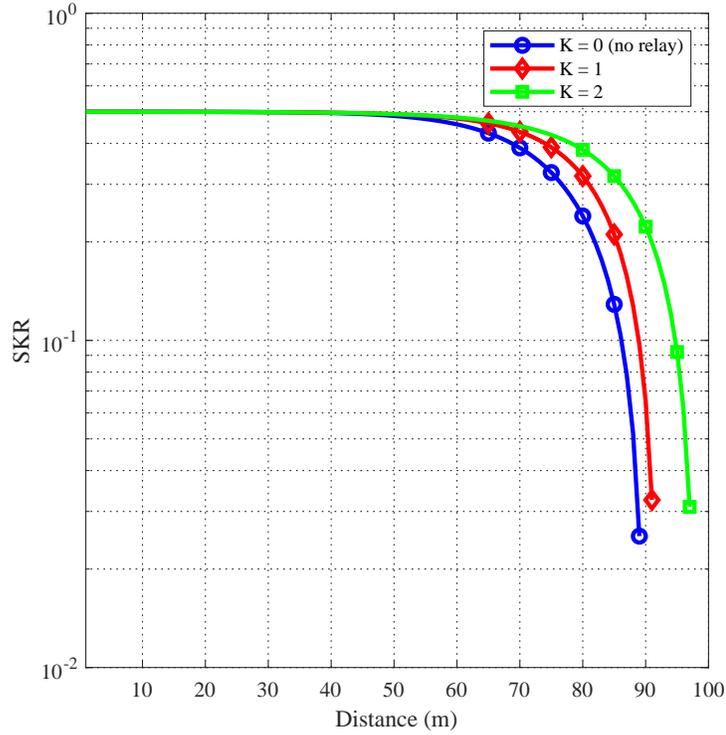}
\caption{SKR of the relay-assisted QKD system over clear ocean with strong turbulence conditions.}
\label{fig:p_2_3}
\end{figure}

The aforementioned end-to-end distances are achievable under the assumption of perfect error correction, i.e., $f = 1$. In an effort to have an insight into what end-to-end distances can be achieved with a practical coding scheme, Fig. \ref{fig:p_2_3} depicts the SKR performance for clear ocean with strong turbulence conditions. We employ an LDPC code with a rate of ${R_c} = 0.5$ optimized for a BSC channel with crossover probability of ${\rm{QBE}}{{\rm{R}}_{th}} = 0.1071 \approx 0.11$ \cite{p1_40}. Although it is possible to use other LDPC codes in \cite{p1_40} optimized for lower QBER values to improve SKR, the maximum achievable distance will still remain the same because the highest QBER that can be tolerated to obtain non-zero SKR should be less than $0.11$. In other words, the maximum achievable distances observed through SKR analysis remain almost the same as those obtained through QBER analysis.

\begin{figure}[t]
\centering
\includegraphics[width=0.7\textwidth]{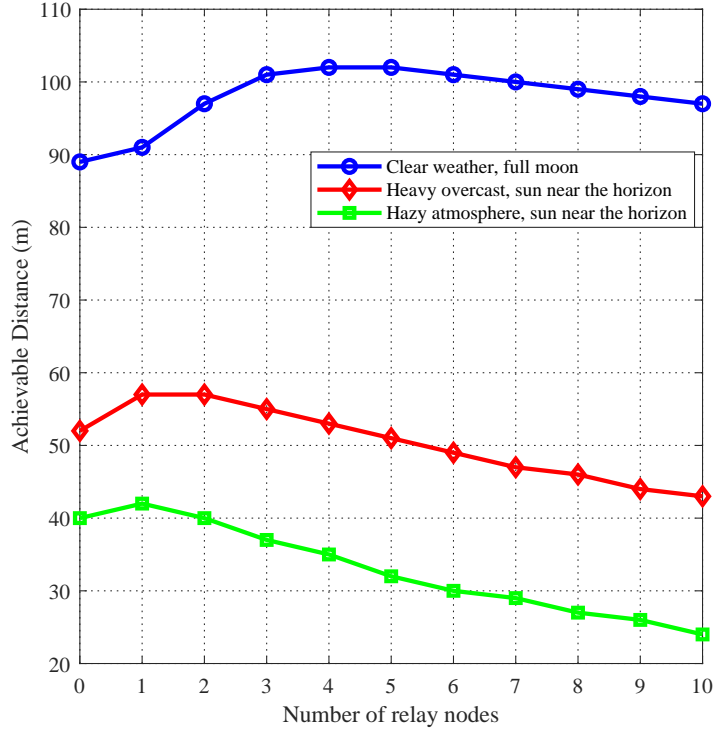}
\caption{Achievable distance versus the number of relay nodes for different atmospheric conditions.}
\label{fig:p_2_4}
\end{figure}

In Fig. \ref{fig:p_2_4}, we investigate the effect of atmospheric conditions on the achievable distance for different number of relay nodes. We consider clear ocean with strong turbulence and assume both hazy and heavy overcast atmosphere when sun is near the horizon at day time. As a benchmark, clear atmospheric conditions at night with a full moon (assumed in Fig. \ref{fig:p_2_2}) is also included. It is observed that as the environment irradiance increases the optimal number of relays (in the sense of maximizing the achievable distance) decreases. Specifically, the maximum achievable distance for heavy overcast and hazy atmosphere are respectively $57$ m and $42$ m when we employ one relay node. These are much lower than $102$ m achievable with four relay nodes at night time.

\begin{figure}[t]
\centering
\includegraphics[width=0.7\textwidth]{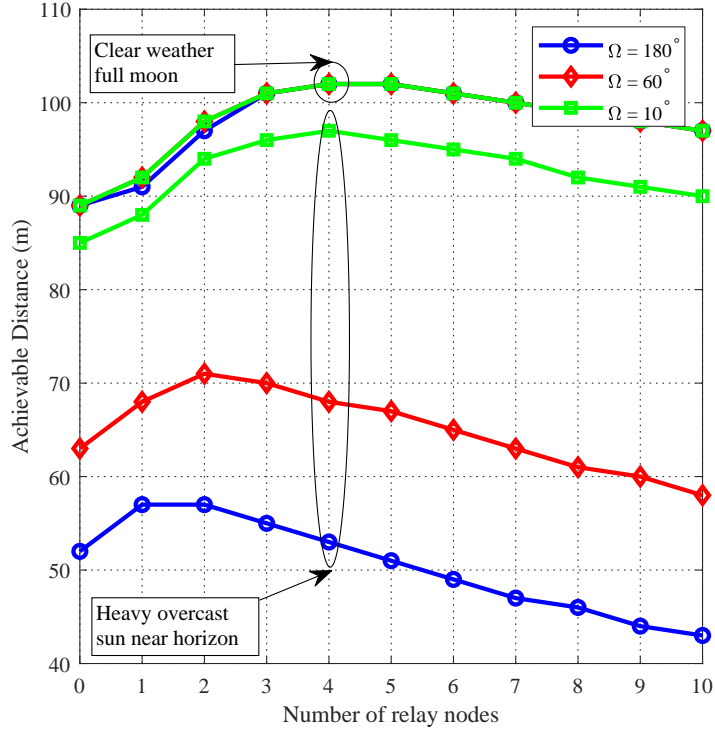}
\caption{Achievable distance versus the number of relay nodes for different FOV values.}
\label{fig:p_2_5}
\end{figure}

In Fig. \ref{fig:p_2_5}, we investigate the effect of FOV on the achievable distance for different number of relay nodes. As atmospheric conditions, we assume clear weather with full moon and heavy overcast. We consider three different FOV values, i.e., $\Omega = {10^ \circ }$, ${60^ \circ }$, and ${180^ \circ }$. It is observed that at night, the achievable distances are almost identical and independent of FOV values, i.e., all three plots overlap with each other. The maximum achievable distance is obtained as $102$ m when we employ four relay nodes. However, further increase in relay nodes does not improve the performance since, according to \eqref{eq:p2_11}, increasing the number of relay nodes leads to an increase in the background noise redirected from relays to Bob's receiver.

Benefit of choosing a proper value of FOV becomes clear as the environment irradiance increases, see plots associated with day time (i.e., heavy overcast). Our results demonstrate that the optimal number of relays (in the sense of maximizing the achievable distance) increases as the FOV decreases. Specifically, maximum achievable distance for $\Omega = {180^ \circ }$ is $57$ m which can be attained by employing one relay. On the other hand, the optimal number of relays for $\Omega = {60^ \circ }$ and $\Omega = {10^ \circ }$ to attain the maximum achievable distance is two and four relay nodes, respectively. As can be readily checked from \eqref{eq:p2_10}, the effect of FOV on ${n_{B0}}$ is more pronounced at day time due to higher value of environment irradiance. Thus, increasing FOV results in increase of the background noise at each relay node and consequently, this increases the background noise redirected from relays to Bob's receiver.

\begin{figure}[t]
\centering
\includegraphics[width=0.7\textwidth]{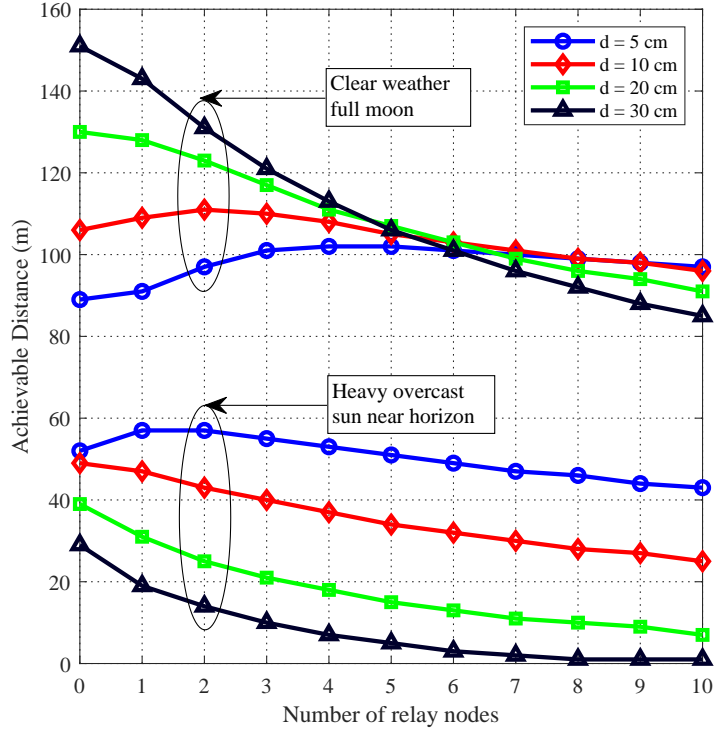}
\caption{Achievable distance versus the number of relay nodes for different aperture sizes.}
\label{fig:p_2_6}
\end{figure}

In Fig. \ref{fig:p_2_6}, we investigate the effect of aperture size on the achievable distance for different number of relay nodes. 
During night time, it is observed that the direct transmission (i.e., $K = 0$) with the largest diameter size under consideration ($d = 30$ cm) achieves the largest transmission distance. For $d = 30$ cm, relay-assisted scheme does not bring any improvement. Actually, its performance is even worse than direct transmission. This is as a result of the accumulation of background noise redirected from relays to Bob for such a large diameter size. On the other hand, when the background noise is limited via the selection of a smaller diameter (i.e., $d = 5$ and $10$ cm), relaying can improve the achievable distance to a certain extent if the relay number is sufficiently small. For example, the maximum achievable distance for $d = 10$ cm is $111$ m which is achieved by employing $K = 2$ relays. If the relay number gets larger (i.e., $K > 2$) for $d = 10$ cm, the accumulated noise becomes too large and the achievable distance decreases. During day time, it is observed that the smallest diameter size under consideration (i.e., $d = 5$ cm) yields larger achievable distances in comparison to other diameter sizes. The maximum achievable distance for $d = 5$ cm is 57 m which is achieved by employing $K = 1$ relay. On the other hand, relaying fails to improve the achievable distance for larger diameters.

As observed from the results presented in Figs. {\ref{fig:p_2_4}}, {\ref{fig:p_2_5}} and {\ref{fig:p_2_6}}, there is a trade-off among system parameters. To better emphasize this, we present Table {\ref{table_p2_2}} which provides the combination of \say{maximum achievable distance} and \say{required number of relay nodes to achieve that distance} for a given set of system parameters and weather conditions. For example, consider a diameter size of $d = 5$ cm, FOV of  $\Omega  = {10^ \circ }$ and heavy overcast. For these given channel and system parameters, the maximum achievable distance is 97 m and this is possible with $K = 4$ relays. It can be readily checked from Table {\ref{table_p2_2}} that the achievable distance is 98 m for a QKD system with $d = 10$ cm, $\Omega  = {10^ \circ }$, and $K = 2$ relays under the same weather conditions. As an another example, consider $d = 20$ cm, $\Omega  = {10^ \circ }$ and hazy overcast. For these given parameters, the maximum achievable distance is 82 m and this is possible with direct transmission (i.e., no relay). For the same weather conditions, QKD systems with parameter sets of $\{d = 10$ cm, $\Omega  = {10^ \circ }$, $K = 1\}$ and $\{d = 5$ cm, $\Omega  = {10^ \circ }$, $K = 3\}$ respectively achieve 84 m and 87 m. Such observations indicate that similar achievable distances can be obtained for different combinations of system parameters. The final selection can be made by the system designer taking into account cost of related equipment, i.e., more relays or larger aperture etc.

\begin{table}
\captionsetup{justification=centering}
\caption{The maximum achievable distance and the required number of relay nodes to achieve that distance for different combinations of system parameters}
\label{table_p2_2}
\begin{center}
\scalebox{0.65}{
\begin{tabular}{|l|c|c|c|c|c|c|c|c|c|c|c|c|}
\hline
\textbf{Diameter size ($d$)} & \multicolumn{3}{c|}{5 cm} & \multicolumn{3}{c|}{10 cm} & \multicolumn{3}{c|}{20 cm} & \multicolumn{3}{c|}{30 cm} \\ \hline
\textbf{Field of view ($\Omega$)} & $10^\circ$ & $60^\circ$ & $180^\circ$ & $10^\circ$ & $60^\circ$ & $180^\circ$ & $10^\circ$ & $60^\circ$ & $180^\circ$ & $10^\circ$ & $60^\circ$ & $180^\circ$ \\ \hline
\textbf{\begin{tabular}[c]{@{}l@{}}Clear weather \\ full moon\end{tabular}} & \begin{tabular}[c]{@{}c@{}}102 m\\ $K$ = 4\end{tabular} & \begin{tabular}[c]{@{}c@{}}102 m\\ $K$ = 4\end{tabular} & \begin{tabular}[c]{@{}c@{}}102 m\\ $K$ = 4\end{tabular} & \begin{tabular}[c]{@{}c@{}}112 m\\ $K$ = 2\end{tabular} & \begin{tabular}[c]{@{}c@{}}112 m\\ $K$ = 2\end{tabular} & \begin{tabular}[c]{@{}c@{}}111 m\\ $K$ = 2\end{tabular} & \begin{tabular}[c]{@{}c@{}}131 m\\ $K$ = 0\end{tabular} & \begin{tabular}[c]{@{}c@{}}131 m\\ $K$ = 0\end{tabular} & \begin{tabular}[c]{@{}c@{}}130 m\\ $K$ = 0\end{tabular} & \begin{tabular}[c]{@{}c@{}}156 m\\ $K$ = 0\end{tabular} & \begin{tabular}[c]{@{}c@{}}154 m\\ $K$ = 0\end{tabular} & \begin{tabular}[c]{@{}c@{}}151 m\\ $K$ = 0\end{tabular} \\ \hline
\textbf{\begin{tabular}[c]{@{}l@{}}Heavy overcast \\ sun near horizon\end{tabular}} & \begin{tabular}[c]{@{}c@{}}97 m\\ $K$ = 4\end{tabular} & \begin{tabular}[c]{@{}c@{}}71 m\\ $K$ = 2\end{tabular} & \begin{tabular}[c]{@{}c@{}}57 m\\ $K$ = 1\end{tabular} & \begin{tabular}[c]{@{}c@{}}98 m\\ $K$ = 2\end{tabular} & \begin{tabular}[c]{@{}c@{}}64 m\\ $K$ = 1\end{tabular} & \begin{tabular}[c]{@{}c@{}}49 m\\ $K$ = 0\end{tabular} & \begin{tabular}[c]{@{}c@{}}100 m\\ $K$ = 0\end{tabular} & \begin{tabular}[c]{@{}c@{}}57 m\\ $K$ = 0\end{tabular} & \begin{tabular}[c]{@{}c@{}}38 m\\ $K$ = 0\end{tabular} & \begin{tabular}[c]{@{}c@{}}106 m\\ $K$ = 0\end{tabular} & \begin{tabular}[c]{@{}c@{}}51 m\\ $K$ = 0\end{tabular} & \begin{tabular}[c]{@{}c@{}}27 m\\ $K$ = 0\end{tabular} \\ \hline
\textbf{\begin{tabular}[c]{@{}l@{}}Hazy overcast \\ sun near horizon\end{tabular}} & \begin{tabular}[c]{@{}c@{}}87 m\\ $K$ =3\end{tabular} & \begin{tabular}[c]{@{}c@{}}55 m\\ $K$ = 1\end{tabular} & \begin{tabular}[c]{@{}c@{}}42 m\\ $K$ =1\end{tabular} & \begin{tabular}[c]{@{}c@{}}84 m\\ $K$ = 1\end{tabular} & \begin{tabular}[c]{@{}c@{}}47 m\\ $K$ = 0\end{tabular} & \begin{tabular}[c]{@{}c@{}}32 m\\ $K$ = 0\end{tabular} & \begin{tabular}[c]{@{}c@{}}81 m\\ $K$ = 0\end{tabular} & \begin{tabular}[c]{@{}c@{}}35 m\\ $K$ = 0\end{tabular} & \begin{tabular}[c]{@{}c@{}}17 m\\ $K$ = 0\end{tabular} & \begin{tabular}[c]{@{}c@{}}82 m\\ $K$ = 0\end{tabular} & \begin{tabular}[c]{@{}c@{}}25 m\\ $K$ = 0\end{tabular} & \begin{tabular}[c]{@{}c@{}}7 m\\ $K$ = 0\end{tabular} \\ \hline
\end{tabular}
}
\end{center}
\end{table}

\begin{figure}
\centering
\includegraphics[width=0.7\textwidth]{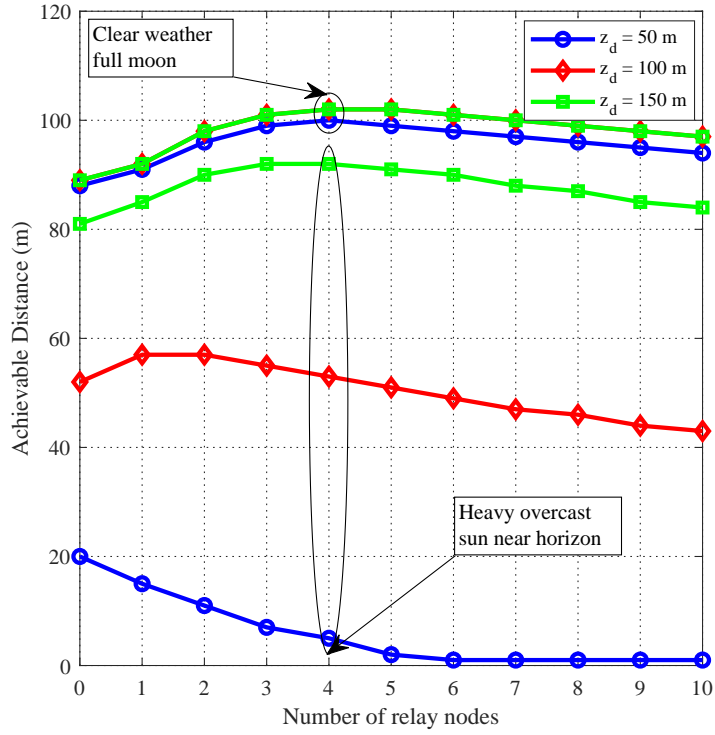}
\caption{Achievable distance versus the number of relay nodes for different depth.}
\label{fig:p_2_7}
\end{figure}

In Fig. {\ref{fig:p_2_7}}, we investigate the effect of depth on the achievable distance for different number of relay nodes. It can be observed that at night time the effect of depth is practically negligible, and the achievable distance remains almost the same for all depths under consideration. The effect of depth becomes more pronounced as the environment irradiance increases. During day time, as the depth increases the optimal number of relay nodes (in the sense of maximizing the achievable distance) increases. Specifically, the maximum achievable distance for depth of ${z_d} = 50$ m is 20 m which is feasible with direct transmission ($K = 0$) and relaying is not required. However, as the depth increases, relaying might be beneficial as a result of decrease in the refracted sunlight from the surface of the water. Specifically, the maximum achievable distances for  ${z_d} = 100$ m and  ${z_d} = 150$ m are respectively 57 m and 92 m. These are achieved respectively by using $K = 1$ and $K = 3$ relays.

\chapter{Decoy State QKD over Underwater Turbulence Channels}\label{ch:decoy}

Real-life QKD systems are often based on attenuated laser pulses i.e., weak coherent states, which occasionally give out more than one photon. This opens up the possibility of sophisticated eavesdropping attacks such as a photon number splitting attack, where Eve stops all single-photon signals and splits multiphoton signals, keeping one copy herself and resending the rest to Bob. Fortunately, the decoy method is able to combat such a sophisticated attack with utilizing different pulse intensities of decoy states dispersed randomly within the signal pulses \cite{p4_14}. The decoy state concept was first introduced by Hwang \cite{p4_15} and the first complete security proof of the decoy-method was given in 2005 by Lo \textit{et al.} \cite{p4_16} considering an infinite amount of intensities. The main idea behind decoy protocols is that the transmitter sends the decoy state pulse sequence (contains no useful information) accompanying the single photon pulse sequence. Since Eve cannot distinguish whether a photon state is from a signal or a decoy, her attempts on photon number splitting attack leads to a variation on the expected yield of signal and decoy states. The decoy QKD protocols were earlier studied for atmospheric, fiber and satellite links \cite{p4_17, p4_18, p4_19}, however those results are not directly applicable to underwater environments with different channel characteristics. There have been only some sporadic efforts on investigating the performance of underwater decoy QKD \cite{p1_19, p4_21, p4_22, p3_12, p3_11}. Specifically, in \cite{p1_19}, the performance of underwater free-space QKD with optimized decoy state protocol was studied based on the Monte Carlo-based channel modeling.  In \cite{p4_21}, Li \textit{et al.} successfully implemented an underwater decoy QKD system through 0.5 m seawater in the laboratory. The experiment in \cite{p3_12} was conducted in a 10 m water tank with extinction coefficient of 0.08 $m^{-1}$. In another implementation \cite{p3_11}, the key generation rate for decoy QKD over an underwater channel of 30 m was studied.

 The current works on underwater decoy QKD only consider the path loss and neglect the effect of underwater turbulence. The effect of underwater turbulence on decoy QKD was only addressed in \cite{p4_22}. In this implementation, Hufnagel \textit{et al.} demonstrate the feasibility of underwater decoy QKD over a 30 m flume tank.

In this chapter, we investigate the fundamental performance limits of decoy BB84 protocol over turbulent underwater channels and provide a comprehensive performance characterization. As path loss model, we consider a modified version of Beer-Lambert formula, which takes into account the effect of scattering. Based on near field analysis \cite{p2_14}, we utilize the wave structure function to determine the average power transfer over turbulent underwater path and use this to obtain a lower bound on key generation rate. Based on this bound, we present the performance of decoy BB84 protocol in different water types (clear and coastal). We further investigate the effect of transmit aperture size and detector field of view (FOV).

The remainder of this chapter is organized as follows. In Section \ref{P4_sys}, we describe system model based on decoy state BB84 QKD protocol. In Section \ref{P4_per}, we derive a lower bound on the key generation rate of the system. In Section \ref{P4_sim}, we present numerical results.

\section{System Model}\label{P4_sys}
\begin{figure}[tb]
\centering
\includegraphics[width=0.75\linewidth]{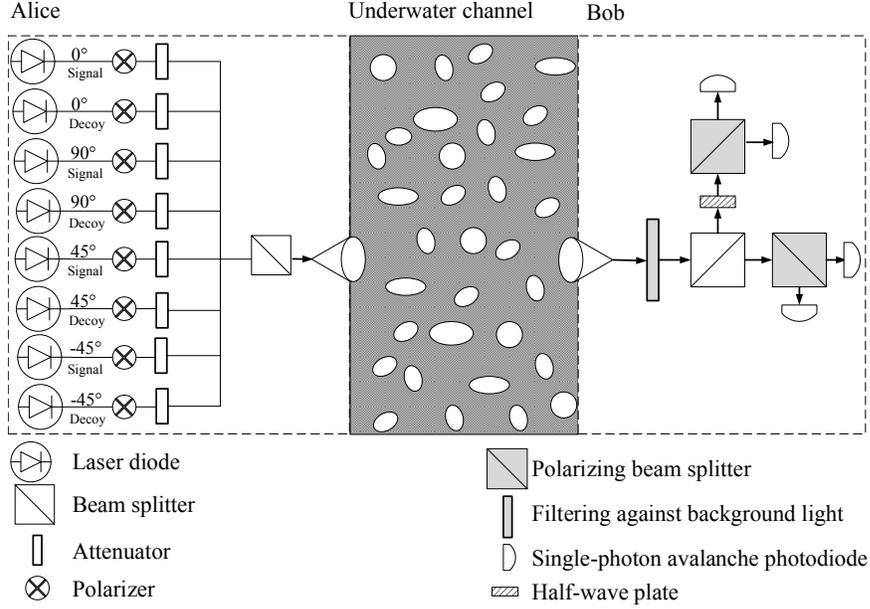}
\caption{Underwater decoy BB84 QKD system under consideration}
\label{fig:p4_1}
\end{figure}

Fig. \ref{fig:p4_1} illustrates a schematic diagram of the decoy QKD system under consideration\footnote{It has been shown that two decoy states are sufficient for practical use and provides the same capability to detect an eavesdropper as an infinite number of decoy states \cite{p4_14}}. The QKD system is built upon two decoy states (i.e., vacuum and weak decoy state) BB84 protocol \cite{p4_14} which aims to create a secret key between the authorized parties (Alice and Bob) such that eavesdropper (Eve) fails to acquire meaningful information. This protocol is based on the principle of polarization encoding. In this protocol, Alice prepares a qubit by choosing the intensity and the basis to encode her bit. She selects from three possible intensities, namely, vacuum state, weak decoy, and signal state. The photon number of each pulse follows a Poisson distribution with the expected photon number of $\mu $ for signal and $v$ ($v < \mu $) for weak decoy.

The basis is chosen randomly between two linear polarization bases namely rectilinear (denoted by $ \oplus $) or diagonal (denoted by $ \otimes $) for every bit she wants to send. She selects a random bit value \say{0} or \say{1} and uses polarization encoding of photons where polarization of ${{0^\circ } \mathord{\left/
 {\vphantom {{0^\circ } { - 45^\circ }}} \right.
 \kern-\nulldelimiterspace} { - 45^\circ }}$ represents 0 and polarization of ${{ + 90^\circ } \mathord{\left/
 {\vphantom {{ + 90^\circ } { + 45^\circ }}} \right.
 \kern-\nulldelimiterspace} { + 45^\circ }}$ represents 1. Through vacuum decoy state, Alice switches off her photon source. At the receiver side, Bob measures the arriving photon randomly in either $ \oplus $ or $ \otimes $ bases. Alice announces the basis and signal/decoy information she has used, while Bob announces the locations of valid detections and the bases used for his measurement. If Alice and Bob have chosen the same basis, they keep the corresponding bits as the sifted key. Alice and Bob can recognize the sift events by transferring information over an authenticated channel communication channel (underwater optical channel in our case). Based on the sifted qubits, a shared one-time pad key is created to use for secure communication \cite{p1_26}. 

Alice uses a weak coherent state laser source with a circular exit pupil and a diameter of ${d_1}$ as the transmitter. Bob collects the light received from Alice with a diameter of ${d_2}$ in the $z = l$ plane. The effects of diffraction, turbulence, and attenuation loss lead to a reduction in Bob's collected photons. In addition, Bob's receiver will collect ${n_B}$ background photons per polarization on average, and each of his detectors will be subject to an average equivalent dark current photon number of ${n_D}$. By considering the dark current and irradiance of the environment, the total average number of noise photons reaching all four Bob's detector can be expressed \cite{p2_14, p1_27}

 \begin{equation}\label{eq:p4_1}
 {Y_0} = 4\left( {{{{n_B}} \mathord{\left/
 {\vphantom {{{n_B}} 2}} \right.
 \kern-\nulldelimiterspace} 2} + {n_D}} \right) = 4{I_{dc}}\Delta t + \frac{{\pi {R_d}A\Delta t'\lambda \Delta \lambda \left( {1 - \cos \left( \Omega  \right)} \right)}}{{{h_p}{c_{light}}}}
 \end{equation}
                                                        
where ${I_{dc}}$ is the dark current, $A$ is the receiver aperture area, $\Omega $ is the field of view of the detector, ${h_p}$ is Planck's constant, ${c_{light}}$ is the speed of light, ${R_d}$ is the irradiance of the environment, $\Delta \lambda $ is the filter spectral width, $\Delta t$ is the bit period and $\Delta t'$ is the receiver gate time. It is convenient to write the depth dependence of ${R_d}\left( {\lambda ,{z_d}} \right)$ as

 \begin{equation}\label{eq:p4_2}
 {R_d}\left( {\lambda ,{z_d}} \right) = {R_d}\left( {\lambda ,0} \right){e^{ - {K_\infty }{z_d}}}
 \end{equation}                                                                                                       
where ${K_\infty }$ is the asymptotic value of the spectral diffuse attenuation coefficient for spectral downwelling plane irradiance \cite{p1_28}. 

\section{Performance Analysis}\label{P4_per}
We define the key generation rate as the ratio of the final key length to the total number of pulses sent by Alice. 
The lower bound on the final key generation rate for QKD system with two decoy states (i.e., vacuum and weak) can be expressed as \cite{p4_14}

 \begin{equation}\label{eq:p4_3}
 R \ge q\left\{ { - {Q_\mu }f\left( {{E_\mu }} \right){H_2}\left( {{E_\mu }} \right) + Q_1^{L,\nu ,0}\left[ {1 - {H_2}\left( {e_1^{U,\nu ,0}} \right)} \right]} \right\}
 \end{equation}
 
where  $q = {1 \mathord{\left/
 {\vphantom {1 2}} \right.
 \kern-\nulldelimiterspace} 2}$ is the basis reconciliation factor, ${Q_\mu }$ is the gain of the signal states, $Q_1^{L,v,0}$ is the lower bound on the gain of single photon state, ${E_\mu }$ is the overall QBER, $e_1^{U,v,0}$ is the error rate of single photon state, $f\left( x \right)$ is the bidirectional error correction efficiency, and ${H_2}\left( x \right) =  - x{\log _2}\left( x \right) - \left( {1 - x} \right){\log _2}\left( {1 - x} \right)$ is the binary Shannon information function. 
 
In \eqref{eq:p4_3}, $Q_1^{L,v,0}$, and $e_1^{U,v,0}$ are respectively defined by \cite{p4_14}

 \begin{equation}\label{eq:p4_4}
 Q_1^{L,v,0} = \frac{{{\mu ^2}{e^{ - \mu }}}}{{\mu \nu  - {\nu ^2}}}\left( {{Q_\nu }{e^\nu } - {Q_\mu }{e^\mu }\frac{{{\nu ^2}}}{{{\mu ^2}}} - \frac{{{\mu ^2} - {\nu ^2}}}{{{\mu ^2}}}{Y_0}} \right)
 \end{equation}
 
  \begin{equation}\label{eq:p4_5}
  e_1^{U,\nu ,0} = \frac{{{E_\nu }{Q_\nu }{e^\nu } - {e_0}{Y_0}}}{{Y_1^{L,\nu ,0}\nu }}
  \end{equation}
where the lower bound on the yield of single photon state (i.e., $Y_1^{L,\nu ,0}$) can be expressed as \cite{p4_14}

  \begin{equation}\label{eq:p4_6}
  Y_1^{L,v,0} = \frac{\mu }{{\mu \nu  - {\nu ^2}}}\left( {{Q_\nu }{e^\nu } - {Q_\mu }{e^\mu }\frac{{{\nu ^2}}}{{{\mu ^2}}} - \frac{{{\mu ^2} - {\nu ^2}}}{{{\mu ^2}}}{Y_0}} \right) 
  \end{equation}
In addition, the overall gain and overall QBER can be respectively expressed as \cite{p4_14}

  \begin{equation}\label{eq:p4_7}
  {Q_\mu } = {Y_0} + 1 - {e^{ - \eta \mu }}
  \end{equation}
  
    \begin{equation}\label{eq:p4_8}
    {E_\mu }{Q_\mu } = {e_0}{Y_0} + {e_{\det }}\left( {1 - {e^{ - \eta \mu }}} \right)
    \end{equation}
                                                                                                       
where $\eta $ denotes the received fraction of transmitted photons, ${e_0} = 0.5$ is the error rate of noise and ${e_{\det }}$ denotes the probability that a photon hits the erroneous detector. We assume ${e_{\det }}$ is constant and independent of the link distance \cite{p4_14}. 

Calculation of the final key generation rate depends on the operation environment and will be discussed in the following for the underwater channel under consideration. To find the performance bounds for the decoy state QKD, it is required to find the fraction of received photons owing to the effect of path loss and turbulence experienced in underwater channels. 

Let $h\left( l \right)$ denote the deterministic path loss depending on the scattering and absorption. Furthermore, let $0 \le \mathord{\buildrel{\lower3pt\hbox{$\scriptscriptstyle\frown$}} 
\over \alpha }  \le 1$ denote the so-called \say{power transfer} term \cite{p1_31} which defines the probability of transmitted photon being reliably received in the presence of turbulence. Therefore, we can write the received fraction of transmitted photons as $\eta  = h\left( l \right)\mathord{\buildrel{\lower3pt\hbox{$\scriptscriptstyle\frown$}} 
\over \alpha } {\eta _{{\rm{Bob}}}}$ where ${\eta _{{\rm{Bob}}}}$ is the transmittance in Bob's side \cite{p4_14, p1_31}. 

For collimated light sources, the geometrical loss is negligible; therefore, the path loss for a laser diode transmitter only depends on the attenuation loss including the effects of absorption and scattering. Based on the modified version of Beer-Lambert formula \cite{p1_33}, the underwater path loss can be expressed as

  \begin{equation}\label{eq:p4_9}
  h\left( l \right) = \exp \left[ { - \varsigma l{{\left( {\frac{{{d_2}}}{{\theta l}}} \right)}^T}} \right]
  \end{equation}
                                                           
where $\varsigma$ is extinction coefficient, $l$ is transmission distance, $\theta $ is the full-width transmitter beam divergence angle and $T$ is a correction coefficient based on water type \cite{p1_33}. Extinction coefficient depends on the wavelength and water type. Typical values of extinction coefficients for $\lambda  = 532$ nm (green color) in different water types can be found in \cite{p1_40}.

Finding a statistical description of $\mathord{\buildrel{\lower3pt\hbox{$\scriptscriptstyle\frown$}} 
\over \alpha } $ and, therefore $\eta $, is a formidable task and a closed-form expression is not available in the literature for near field propagation. As an alternative, the lower bound on the average power transfer can be expressed as \cite{p4_14}

 \begin{equation}\label{eq:p4_10}
 {\rm E}\left( {\mathord{\buildrel{\lower3pt\hbox{$\scriptscriptstyle\frown$}} 
\over \alpha } } \right) \ge \alpha  = \frac{{8\sqrt F }}{\pi }\int\limits_0^1 {\exp \left( { - W\left( {dx,l} \right)/2} \right)\left( {{{\cos }^{ - 1}}\left( x \right) - x\sqrt {1 - {x^2}} } \right){J_1}\left( {4x\sqrt F } \right)dx} 
 \end{equation}
                      
where ${J_1}\left(\cdot \right)$ is the first-order Bessel function of the first kind and $F$ is the Fresnel number product of transmit and receive diameters given by $F = {\left( {\pi {d^2}/4\lambda l} \right)^2}$. In \eqref{eq:p4_10}, $W\left(\cdot {,} \cdot\right)$ is the underwater wave structure function. For a given transmission distance of $l$ and given separation distance between two points on the phase front transverse to the axis of propagation (denoted by $\rho $), it is expressed as \cite{p2_14}

\begin{equation}\label{eq:p4_11}
W\left( {\rho ,l} \right) = 1.44\pi {k^2}l\left( {\frac{{\alpha _{{\rm{th}}}^2\chi }}{{{\omega ^2}}}} \right){\varepsilon ^{ - \frac{1}{3}}}\left( {1.175\eta _K^{2/3}\rho  + 0.419{\rho ^{\frac{5}{3}}}} \right)\left( {{\omega ^2} + {d_r} - \omega \left( {{d_r} + 1} \right)} \right)
\end{equation}
                     
where $k = {{2\pi } \mathord{\left/
 {\vphantom {{2\pi } \lambda }} \right.
 \kern-\nulldelimiterspace} \lambda }$ is the wave number, $\omega $ is the relative strength of temperature and salinity fluctuations, $\varepsilon $ is the dissipation rate of turbulent kinetic energy per unit mass of fluid, ${\alpha _{{\rm{th}}}}$ is the thermal expansion coefficient, $\chi $ is the dissipation rate of mean-squared temperature and ${d_r}$ is the eddy diffusivity ratio. In \eqref{eq:p4_11}, ${\eta _K}$ is Kolmogorov microscale length and given by ${\eta _K} = {\left( {{{{\upsilon ^3}} \mathord{\left/
 {\vphantom {{{\upsilon ^3}} \varepsilon }} \right.
 \kern-\nulldelimiterspace} \varepsilon }} \right)^{{1 \mathord{\left/
 {\vphantom {1 4}} \right.
 \kern-\nulldelimiterspace} 4}}}$ with $\upsilon $ referring to the kinematic viscosity. 

To find the performance bounds for the decoy state QKD, assume that $P\left( {\mathord{\buildrel{\lower3pt\hbox{$\scriptscriptstyle\frown$}} 
\over \alpha } } \right)$ denotes the probability density function (pdf) of $\mathord{\buildrel{\lower3pt\hbox{$\scriptscriptstyle\frown$}} 
\over \alpha } $. Although $P\left( {\mathord{\buildrel{\lower3pt\hbox{$\scriptscriptstyle\frown$}} 
\over \alpha } } \right)$ is not available in the literature, here we take the advantage of expected value; i.e., ${\rm E}\left( {\mathord{\buildrel{\lower3pt\hbox{$\scriptscriptstyle\frown$}} 
\over \alpha } } \right) = \int {\mathord{\buildrel{\lower3pt\hbox{$\scriptscriptstyle\frown$}} 
\over \alpha } P\left( {\mathord{\buildrel{\lower3pt\hbox{$\scriptscriptstyle\frown$}} 
\over \alpha } } \right)} d\mathord{\buildrel{\lower3pt\hbox{$\scriptscriptstyle\frown$}} 
\over \alpha }$.

 Since $g\left( x \right) = {e^{ - x}}$ is a convex function with negative derivative, the lower bound on the ${Q_v}$ and ${Q_\mu }$ respectively can be expressed as 
 
\begin{equation}\label{eq:p4_12}
\begin{array}{l}
{Q_\nu } = {Y_0} + 1 - \int\limits_0^1 {{e^{ - \nu {\eta _{Bob}}h\left( l \right)\mathord{\buildrel{\lower3pt\hbox{$\scriptscriptstyle\frown$}} 
\over \alpha } }}} P\left( {\mathord{\buildrel{\lower3pt\hbox{$\scriptscriptstyle\frown$}} 
\over \alpha } } \right)d\mathord{\buildrel{\lower3pt\hbox{$\scriptscriptstyle\frown$}} 
\over \alpha }  \ge {Y_0} + 1 - \left( {\int\limits_0^1 {\left( {1 - \mathord{\buildrel{\lower3pt\hbox{$\scriptscriptstyle\frown$}} 
\over \alpha }  + \mathord{\buildrel{\lower3pt\hbox{$\scriptscriptstyle\frown$}} 
\over \alpha } {e^{ - \nu {\eta _{{\rm{Bob}}}}h\left( l \right)}}} \right)} P\left( {\mathord{\buildrel{\lower3pt\hbox{$\scriptscriptstyle\frown$}} 
\over \alpha } } \right)d\mathord{\buildrel{\lower3pt\hbox{$\scriptscriptstyle\frown$}} 
\over \alpha } } \right)\\
\,\,\,\,\,\, = {Y_0} + 1 - \left( {1 - E\left( {\mathord{\buildrel{\lower3pt\hbox{$\scriptscriptstyle\frown$}} 
\over \alpha } } \right) + E\left( {\mathord{\buildrel{\lower3pt\hbox{$\scriptscriptstyle\frown$}} 
\over \alpha } } \right){e^{ - \nu {\eta _{{\rm{Bob}}}}h\left( l \right)}}} \right)\\
\,\,\,\,\,\, \ge Q_\nu ^L \buildrel \Delta \over = {Y_0} + \alpha  - \alpha \exp \left[ { - \nu {\eta _{{\rm{Bob}}}}h\left( l \right)} \right]
\end{array}
\end{equation}
                           
\begin{equation}\label{eq:p4_13}
{Q_\mu } \ge Q_\mu ^L \buildrel \Delta \over = {Y_0} + \alpha  - \alpha \exp \left[ { - \mu {\eta _{{\rm{Bob}}}}h\left( l \right)} \right]
\end{equation}
                                                                                  
Ignoring the effects of diffraction and turbulence, the upper bound on ${Q_\mu }$ can be calculated by inserting $\mathord{\buildrel{\lower3pt\hbox{$\scriptscriptstyle\frown$}} 
\over \alpha }  = 1$ into \eqref{eq:p4_7} and can be written as

\begin{equation}\label{eq:p4_14}
{\left. {{Q_\mu }} \right|_{\mathord{\buildrel{\lower3pt\hbox{$\scriptscriptstyle\frown$}} 
\over \alpha }  = 1}} = Q_\mu ^U \buildrel \Delta \over = {Y_0} + 1 - \exp \left[ { - \mu {\eta _{{\rm{Bob}}}}h\left( l \right)} \right]
\end{equation}
                                                                                 
Therefore, the upper bound on the overall QBER can be expressed as

\begin{equation}\label{eq:p4_15}
E_\mu ^U \buildrel \Delta \over = \frac{{{e_0}{Y_0} + {e_{\det }}\left( {1 - {e^{ - \mu {\eta _{{\rm{Bob}}}}h\left( l \right)}}} \right)}}{{Q_\mu ^L}}
\end{equation}
                                                                                                 
Employing \eqref{eq:p4_12}, \eqref{eq:p4_13}, \eqref{eq:p4_14}, and \eqref{eq:p4_15}, the lower bound on $Y_1^{L,v,0}$ and $Q_1^{L,v,0}$, and the upper bound on $e_1^{U,v,0}$ respectively can be written as 

\begin{equation}\label{eq:p4_16}
Y_1^{L,\nu ,0}\, \ge Y_1^L\, \buildrel \Delta \over = \frac{\mu }{{\mu \nu  - {\nu ^2}}}\left( {Q_\nu ^L{e^\nu } - Q_\mu ^U{e^\mu }\frac{{{\nu ^2}}}{{{\mu ^2}}} - \frac{{{\mu ^2} - {\nu ^2}}}{{{\mu ^2}}}{Y_0}} \right)
\end{equation}

\begin{equation}\label{eq:p4_17}
Q_1^{L,\nu ,0} \ge Q_1^L \buildrel \Delta \over = \frac{{{\mu ^2}{e^{ - \mu }}}}{{\mu \nu  - {\nu ^2}}}\left( {Q_\nu ^L{e^\nu } - Q_\mu ^U{e^\mu }\frac{{{\nu ^2}}}{{{\mu ^2}}} - \frac{{{\mu ^2} - {\nu ^2}}}{{{\mu ^2}}}{Y_0}} \right)
\end{equation}

\begin{equation}\label{eq:p4_18}
e_1^{U,\nu ,0} \le e_1^U \buildrel \Delta \over = \frac{{\left( {{e_0}{Y_0} + {e_{\det }}\left( {1 - {e^{ - v{\eta _{\det }}h\left( l \right)}}} \right)} \right){e^\nu } - {e_0}{Y_0}}}{{Y_1^L\nu }}
\end{equation}
                                                                       
Thus, the final expression for the lower bound on key generation rate can be calculated as

\begin{equation}\label{eq:p4_19}
{R^L} = q\left\{ { - Q_\mu ^Uf\left( {E_\mu ^U} \right){H_2}\left( {E_\mu ^U} \right) + Q_1^L\left[ {1 - {H_2}\left( {e_1^U} \right)} \right]} \right\}
\end{equation}

\textbf{Special case:} As a benchmark, we consider a QKD system which uses BB84 protocol for key distribution. By considering an \say{ideal} single-photon transmitter, the gain of the signal states and overall QBER will be equal to the gain of single photon state and error rate of single photon state, respectively; i.e., ${Q_\mu } = {Q_1}$ and ${E_\mu } = {e_1}$. By substituting these parameters in \eqref{eq:p4_3} and after some simple mathematical manipulations, the lower bound on the final key generation rate can be expressed as

\begin{equation}\label{eq:p4_20}
R \ge R_{{\rm{BB84}}}^L \buildrel \Delta \over = \frac{{{Q_1}}}{2}\left\{ {1 - {H_2}\left( {{e_1}} \right)\left[ {1 + f\left( {{e_1}} \right)} \right]} \right\}
\end{equation}
                                                                                  
which the first term (i.e., ${{{Q_1}} \mathord{\left/
 {\vphantom {{{Q_1}} 2}} \right.
 \kern-\nulldelimiterspace} 2}$) indicates the probability of sift and the second term coincides with Eq. \eqref{eq:p4_3} in \cite{p1_40}. Ignoring the turbulence effect and recalling Eqs. (9) and (10) from \cite{p4_14}, the error rate and single photon gain can be expressed as

\begin{equation}\label{eq:p4_21}
{e_1} = \frac{{\frac{1}{2}{Y_0} + {e_{\det }}\eta }}{{{Y_0} + \eta }}
\end{equation}

\begin{equation}\label{eq:p4_22}
{Q_1} = \left( {{Y_0} + \eta } \right)\mu {e^{ - \mu }}
\end{equation}

Note that when ${e_{\det }} = 0$, \eqref{eq:p4_21} coincides with Eq. (17) in \cite{p3_20}. 

\section{Simulation Results}\label{P4_sim}
In this section, we demonstrate the performance of underwater QKD scheme under consideration. We assume the transmitter beam divergence angle of $\theta  = {6^ \circ }$, the system error of ${e_{\det }} = 3.3\% $, the dark current count rate of ${I_{dc}} = 60$ Hz, filter spectral width of $\Delta \lambda  = 30$ nm, bit period of $\Delta t = 35$ ns, receiver gate time of $\Delta t' = 200$ ps, the transmittance in Bob's side of ${\eta _{Bob}} = 0.045$, the expected photon number of $\mu  = 0.48$ and $v = 0.05$ for signal and decoy state, respectively. Here, we assume the error correction efficiency of 1.22 regardless of the error rate \cite{p4_14}. Unless otherwise stated, we assume the transmitter and receiver aperture diameters of $d = 5$ cm, FOV of $\Omega  = {180^ \circ }$ and clear atmospheric conditions at night with a full moon. The typical total irradiances at sea level, i.e., ${R_d}\left( {\lambda ,0} \right)$, in the visible wavelength band for some typical atmospheric conditions are provided in \cite{p1_29}. As for channel parameters, we assume $\alpha  = 2.56 \times {10^{ - 4}}$  1/deg and $\upsilon  = 1.0576 \times {10^{ - 6}}$  m2s-1 \cite{p1_35}. We consider three representative cases for turbulence strength. Specifically, we assume $\omega = -2.2$, $\chi_T = 2\times10^{-7}$ $\rm{K^{2}s^{-3}}$ and $\varepsilon = 2\times10^{-5}$ $\rm{m^{2}s^{-3}}$  for weak turbulence, $\omega = -2.2$,  $\chi_T = 10^{-6}$ $\rm{K^{2}s^{-3}}$ and $\varepsilon = 5\times10^{-7}$ $\rm{m^{2}s^{-3}}$ for moderate turbulence and  $\omega = -2.2$,  $\chi_T = 10^{-5}$ $\rm{K^{2}s^{-3}}$ and  $\varepsilon = 10^{-5}$ $\rm{m^{2}s^{-3}}$ for strong oceanic turbulence \cite{p1_43}. For the convenience of the reader, the channel and system parameters are summarized in Table \ref{table:p4}.

\begin{table}[t]
\captionsetup{justification=centering}
\caption{System and channel parameters}
\label{table:p4}
\begin{center}
\scalebox{0.9}{
\begin{tabular}{ |l|l|l| } 
 \hline
 \textbf{Parameter} & \textbf{Definition} & \textbf{Numerical Value} \\ \hline
 $\mu $ & \text{Expected photon number for signal} & 0.48 \cite{p4_14} \\ \hline 
 $v$ & \text{Expected photon number for decoy state} & 0.05 \cite{p4_14} \\ \hline 
 ${\eta _{{\rm{Bob}}}}$ & \text{Transmittance in Bob's side} & 0.045\cite{p4_14} \\ \hline 
 ${e_{\det }}$ & \text{System error} & 3.3\% \cite{p4_14} \\ \hline 
$\Omega$ & \text{Field of view} & $180^{\circ}$ \cite{p1_33} \\ \hline 
$\Delta \lambda$ & \text{Filter spectral width} & $30$ $\rm{nm}$ \cite{p2_14}\\ \hline 
$\lambda$ & \text{Wavelength} & $530$ $\rm{nm}$ \cite{p1_33} \\ \hline 
$\Delta t$ & \text{Bit period} & $35$ $\rm{ns}$ \cite{p1_27} \\ \hline 
$\Delta t\ensuremath{'}$ & \text{Receiver gate time}  & $200$ $\rm{ps}$ \cite{p1_27} \\ \hline 
$d_1$ & \text{Transmitter aperture diameter} & $5$ $\rm{cm}$ \cite{p1_31} \\ \hline 
$d_2$ & \text{Receiver aperture diameter} & $5$ $\rm{cm}$ \cite{p1_31} \\ \hline 
$I_{dc}$ & \text{Dark current count rate} & $60$ $\rm{Hz}$ \cite{p1_27} \\ \hline 
$K_{\infty}$ & \text{Asymptotic diffuse attenuation coefficient} & $0.08$ $\rm{m^{-1}}$ \cite{p1_28} \\ \hline 
$z_d$ & \text{Depth} & $100$ $\rm{m}$ \cite{p1_27} \\ \hline 
$\theta$ & \text{Transmitter beam divergence angle} & $6^{\circ}$ \cite{p1_33} \\ \hline 
$\varsigma$ & \text{Extinction coefficient} \,\begin{tabular}{l|l}  & \text{Clear water} \\
 & \text{Coastal water}  \end{tabular} & 
 \begin{tabular}{l} \!\!\!\!$0.151$ $\rm{m^{-1}}$ \cite{p1_32}\\
\!\!\!\! $0.339$ $\rm{m^{-1}}$ \cite{p1_32} \end{tabular} \\\hline
$T$ & \text{Correction coefficient} 
\begin{tabular}{c|c}  
& $\theta  = 6^\circ ,\,\,{d_2} = 5\,{\rm{cm}}$ \,\,\,\,\,\,\,\,\\
 & $\theta  = 6^\circ ,\,\,{d_2} = 10\,{\rm{cm}}$\,\,\,\,\,\,\,\, \\
 & $\theta  = 6^\circ ,\,\,{d_2} = 20\,{\rm{cm}}$ \,\,\,\,\,\,\,\,\end{tabular} & 
 \begin{tabular}{l}  \!\!\!\! $0.13$  \cite{p1_33}\\
\!\!\!\! $0.16$  \cite{p1_33} \\
\!\!\!\! $0.21$  \cite{p1_33} \end{tabular} \\\hline
\end{tabular}}
\end{center}
\end{table}

\begin{figure}[tb]
\centering
\includegraphics[width=0.75\linewidth]{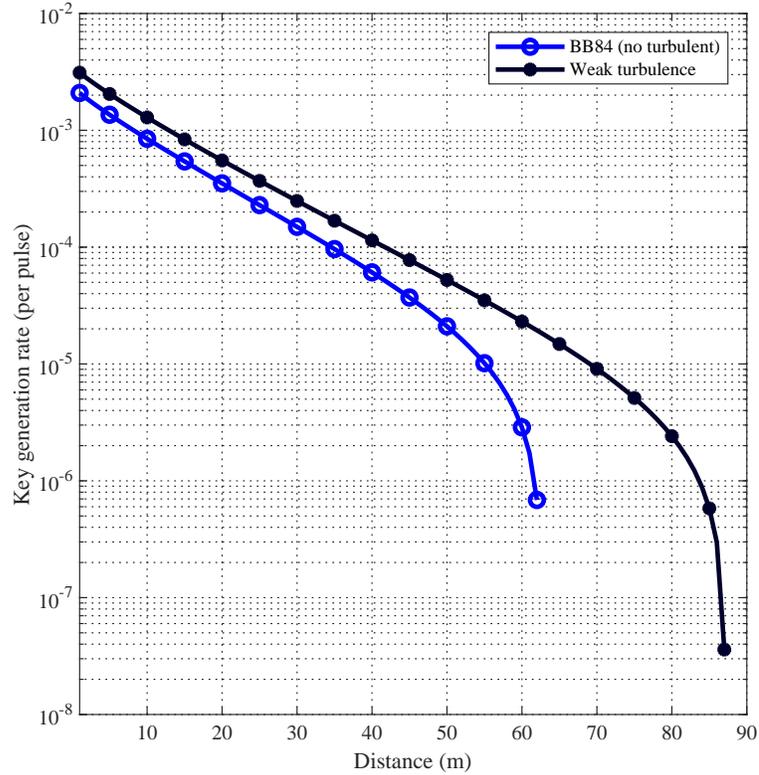}
\caption{Key generation rate (per pulse) of the QKD system over clear ocean for BB84 and decoy BB84 protocols.}
\label{fig:p4_2}
\end{figure}

Fig. \ref{fig:p4_2} illustrates the lower bound on key generation rate with respect to link distance for both BB84 and decoy BB84 protocols. As the water type, we assume clear ocean. We consider decoy BB84 protocol operates over weak turbulence condition, while the BB84 protocol operates over non turbulent condition. As it can be observed from Fig. \ref{fig:p4_2}, there is huge difference between the achievable distances for these two protocols. The achievable distance to obtain non-zero key generation rate for BB84 protocol is around 87 m while it decreases to 62 m for the decoy scheme. Note that we consider an ideal single photon transmitter for BB84 protocol, whereas this condition is not available yet in practice.

\begin{figure}[tb]
\centering
\includegraphics[width=0.75\linewidth]{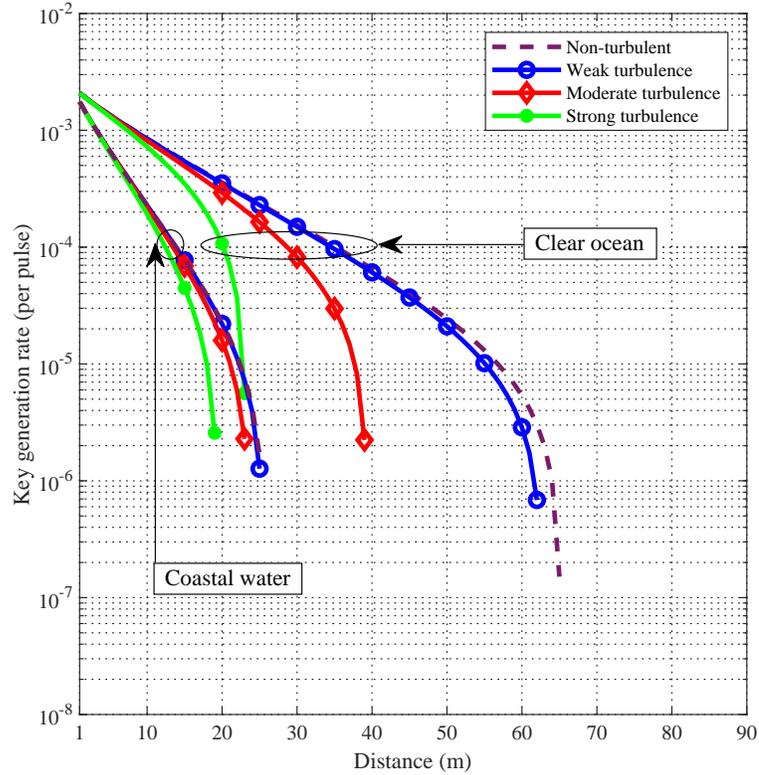}
\caption{Key generation rate (per pulse) of the QKD system over clear ocean and coastal water for non-turbulent, weak, moderate, and strong turbulence conditions.}
\label{fig:p4_3}
\end{figure}

Fig. \ref{fig:p4_3} illustrates the lower bound on key generation rate with respect to link distance for clear ocean, and coastal water. For each water type, we assume weak, moderate and strong turbulence. The lower bound on key generation rate for non-turbulent case is also included as a benchmark. It can be observed from Fig. \ref{fig:p4_3} that the turbulence effect in coastal water is small and the path loss is the dominant factor. The achievable distance to obtain non-zero key generation rate in coastal water with weak turbulence conditions is around 25 m which is the same as non-turbulent case. The achievable distance reduces to 23 m and 19 m for moderate and strong turbulence conditions, respectively. As turbidity decreases, the achievable distance increases and the effect of turbulence is more pronounced. While, the achievable distance to maintain positive key generation rate for clear ocean and non-turbulent conditions is around 65 m, it decreases to 62 m for weak turbulence conditions. The achievable distance reduces to 39 m and 23 m for moderate and strong turbulence, respectively.

\begin{figure}[tb]
\centering
\includegraphics[width=0.75\linewidth]{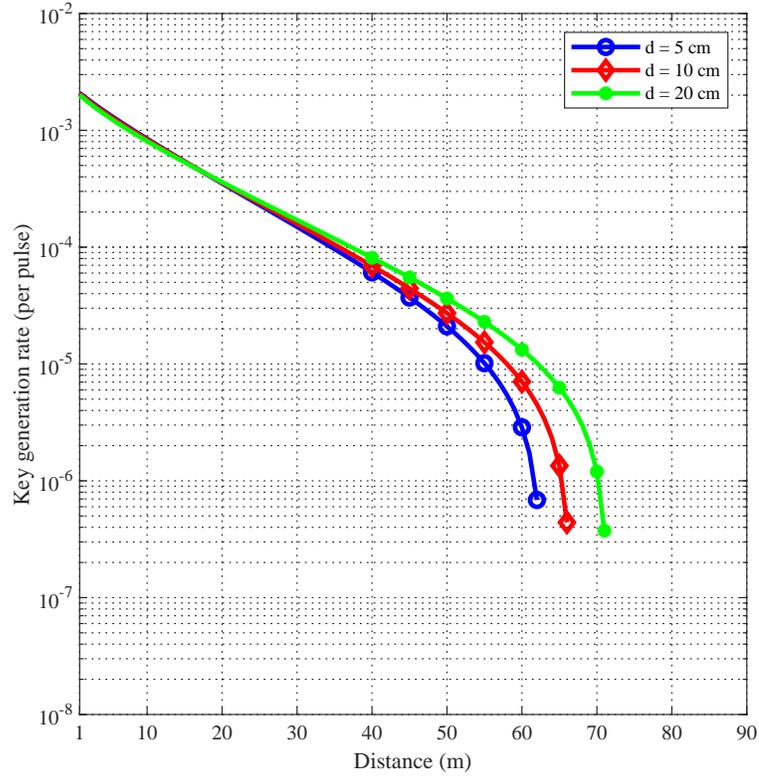}
\caption{Key generation rate (per pulse) of the QKD system over clear ocean with weak turbulence conditions for different diameter size.}
\label{fig:p4_4}
\end{figure}

In Fig. \ref{fig:p4_4}, we study the effect of aperture size on the performance of underwater QKD system. We assume clear ocean with weak turbulence and consider two distinct atmospheric conditions. We assume the receiver aperture size varies as ${d_2} = $ 20 and 30 cm and the transmitter pupil has the same diameter as the receiver. As a benchmark, diameter size of 5 cm (assumed in Fig. \ref{fig:p4_2}) is also included. It is observed that at night time with a full moon, the achievable distance increases as the diameter size increases. For example, the achievable distance for ${d_2} = $ 5 cm is around 62 m, while it climbs up to 66 m and 71 m for ${d_2} = $ 10 cm and 20 cm, respectively. It should be emphasized that the increase in background noise as a result of increasing the diameter size is negligible at night, however, larger diameter results in
an increase of collected photons coming from Alice.

\begin{figure}[tb]
\centering
\includegraphics[width=0.75\linewidth]{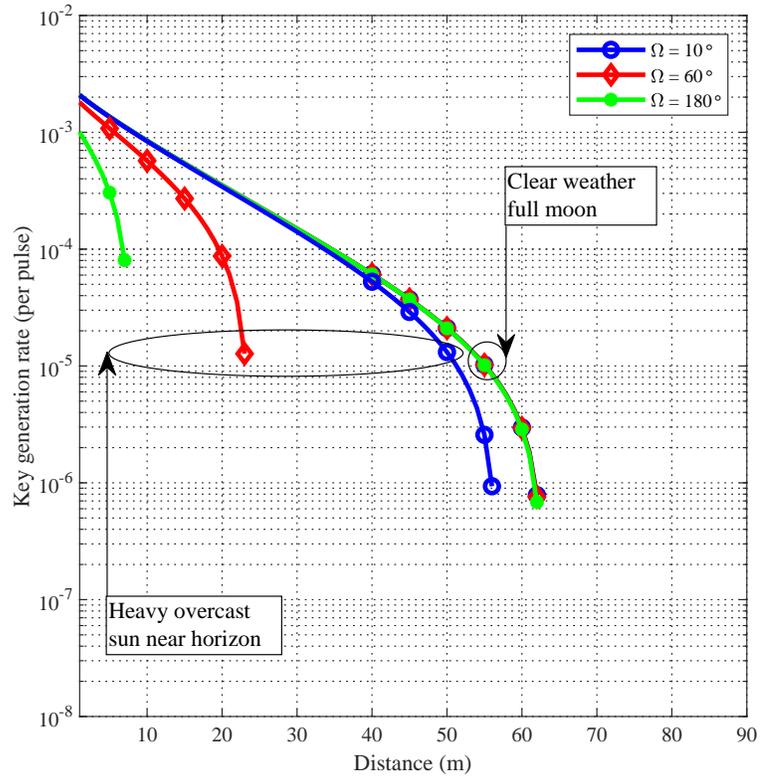}
\caption{Key generation rate (per pulse) of the QKD system over clear ocean with weak turbulence conditions for different FOV.}
\label{fig:p4_5}
\end{figure}

In Fig. \ref{fig:p4_5}, we investigate the effect of FOV on the performance of the QKD system. We assume clear ocean with weak turbulence and consider two atmospheric cases. These are clear weather night with a full moon and heavy overcast when sun is near the horizon. We assume  $\Omega = {10^ \circ }$, ${60^ \circ }$ and ${180^ \circ }$. It is observed that at night time, the effect of FOV is practically negligible and the key generation rate remains the same for all FOV values under consideration. Benefit of choosing a proper value of FOV becomes clear as the environment irradiance increases. In daylight, we observe that the achievable distance significantly improves as the FOV decreases. This improvement is due to the decrease in background noise as the FOV decreases. Mathematically speaking, the achievable distance for $\Omega = {180^ \circ }$  is around 7 m, while it increases to 23 m and 56 m for $\Omega = {60^ \circ }$ and ${10^ \circ }$, respectively.

\chapter{Conclusions}\label{ch:conc}

In Chapter \ref{ch:QKD}, we have investigated the performance of the BB84 protocol over turbulent underwater channels. Our results have demonstrated that the turbulence effect in turbid water is negligible and the path loss is the dominant factor. As turbidity decreases, the achievable distance increases and the effect of turbulence is more pronounced. Our results have further shown that achievable distance for underwater QKD system at day time drastically reduces in comparison to night time due to an increase in the received background noise. We have also investigated the effect of system parameters such as aperture size and FOV on QBER and SKR performance. At night time, the effect of FOV has been found to be practically negligible and the performance remains the same for all FOV values under consideration. In daylight, the achievable distance significantly improves as the FOV, and therefore background noise, decreases. It has been also observed that when the aperture size increases the achievable distance increases at night time while it decreases at daylight. Such observations indicate the necessity of using adaptive selection of aperture size in practical implementations.

In Chapter \ref{ch:relay}, we have investigated the performance of relay assisted underwater QKD with BB84 protocol. Our results have demonstrated that relay-assisted QKD has the potential to increase the end-to-end achievable distance if the system parameters are judiciously selected. While adding relay nodes mitigates the degrading effects of turbulence-induced fading, it also results in an increase of the average number of background photons at Bob's receiver. To investigate this trade-off, we have studied the effect of system parameters such as aperture size and FOV on the achievable distance and determined the optimal number of relays in the sense of maximizing the achievable distance. It is observed that the optimal number of relay increases as the FOV decreases and/or as the receive diameter decreases. Our results highlight that relaying brings improvements when the noise level is kept low (e.g., small receiver diameter, small FOV, and/or low environment irradiance) and the water turbidity is low (e.g., clear ocean). 


In Chapter \ref{ch:decoy}, we have investigated the fundamental performance limits of decoy BB84 protocol over turbulent underwater channels. we utilized the wave structure function to determine the average power transfer over turbulent underwater path and used this to obtain a lower bound on key generation rate. Based on this bound, we presented the performance of decoy BB84 protocol in different water types and compared it with original BB84 protocol.



\begin{postliminary}
\bibliography{example-thesis.bib}


\end{postliminary}

\end{document}